\documentclass[aps, prd, reprint,showpacs,  preprintnumbers,amsmath,amssymb,superscriptaddress, nofootinbib]{revtex4-1}
\usepackage{amssymb, amsmath, braket, nicefrac, graphicx} 
\usepackage[tight]{subfigure}
\usepackage{relsize}
\usepackage{lineno}
\usepackage{setspace}

\usepackage{ esint }
\usepackage{chngpage}
\usepackage{feynmf}
\usepackage{epsfig}
\usepackage{multirow}
\usepackage{rotating}
\usepackage{float}
\usepackage{color}
\usepackage{xcolor}
\usepackage{appendix}
\usepackage{ amstext }
\usepackage{ gensymb }
\usepackage{ textcomp }
\usepackage{ wasysym }

\usepackage{pstricks}

\usepackage{dcolumn}
\usepackage{bm}
\usepackage{units}

\newcommand{\numu}{\mbox{$\nu_{\mu}$}}                   

\newcommand{\anumu}{\mbox{$\overline{\nu}_{\mu}$}}                   
\newcommand{\simgt}{\,\hbox{\lower0.6ex\hbox{$\sim$}\llap{\raise0.6ex\hbox{$>$}}}\,}
\newcommand{\simlt}{\,\hbox{\lower0.6ex\hbox{$\sim$}\llap{\raise0.6ex\hbox{$<$}}}\,}
\definecolor{maroon}{RGB}{162,10,10}

\usepackage{hyperref}
\usepackage[all]{hypcap}
\hypersetup{
    colorlinks,
    citecolor=black,
    filecolor=black,
    linkcolor=black,
    urlcolor=black
}

\makeatletter
\renewenvironment{table}
  {\def\@captype{table}}
  {}

\renewenvironment{figure}
  {\def\@captype{figure}}
  {}
\makeatother
   
\begin{document}
\preprint{FERMILAB-PUB-14-401-E}
\preprint{ (accepted)}

\title{Study of quasielastic scattering using charged-current $\numu$-iron interactions \\in the MINOS Near Detector}       





\newcommand{\Berkeley}{Lawrence Berkeley National Laboratory, Berkeley, California, 94720 USA}
\newcommand{\Cambridge}{Cavendish Laboratory, University of Cambridge, Madingley Road, Cambridge CB3 0HE, United Kingdom}
\newcommand{\Cincinnati}{Department of Physics, University of Cincinnati, Cincinnati, Ohio 45221, USA}
\newcommand{\FNAL}{Fermi National Accelerator Laboratory, Batavia, Illinois 60510, USA}
\newcommand{\RAL}{Rutherford Appleton Laboratory, Science and Technology Facilities Council, Didcot, OX11 0QX, United Kingdom}
\newcommand{\UCL}{Department of Physics and Astronomy, University College London, Gower Street, London WC1E 6BT, United Kingdom}
\newcommand{\Caltech}{Lauritsen Laboratory, California Institute of Technology, Pasadena, California 91125, USA}
\newcommand{\Alabama}{Department of Physics and Astronomy, University of Alabama, Tuscaloosa, Alabama 35487, USA}
\newcommand{\ANL}{Argonne National Laboratory, Argonne, Illinois 60439, USA}
\newcommand{\Benedictine}{Physics Department, Benedictine University, Lisle, Illinois 60532, USA}
\newcommand{\BNL}{Brookhaven National Laboratory, Upton, New York 11973, USA}
\newcommand{\CdF}{APC -- Universit\'{e} Paris 7 Denis Diderot, 10, rue Alice Domon et L\'{e}onie Duquet, F-75205 Paris Cedex 13, France}
\newcommand{\Cleveland}{Cleveland Clinic, Cleveland, Ohio 44195, USA}
\newcommand{\Delhi}{Department of Physics \& Astrophysics, University of Delhi, Delhi 110007, India}
\newcommand{\GEHealth}{GE Healthcare, Florence South Carolina 29501, USA}
\newcommand{\Harvard}{Department of Physics, Harvard University, Cambridge, Massachusetts 02138, USA}
\newcommand{\HolyCross}{Holy Cross College, Notre Dame, Indiana 46556, USA}
\newcommand{\Houston}{Department of Physics, University of Houston, Houston, Texas 77204, USA}
\newcommand{\IIT}{Department of Physics, Illinois Institute of Technology, Chicago, Illinois 60616, USA}
\newcommand{\Iowa}{Department of Physics and Astronomy, Iowa State University, Ames, Iowa 50011 USA}
\newcommand{\Indiana}{Indiana University, Bloomington, Indiana 47405, USA}
\newcommand{\ITEP}{High Energy Experimental Physics Department, ITEP, B. Cheremushkinskaya, 25, 117218 Moscow, Russia}
\newcommand{\JMU}{Physics Department, James Madison University, Harrisonburg, Virginia 22807, USA}
\newcommand{\LASL}{Nuclear Nonproliferation Division, Threat Reduction Directorate, Los Alamos National Laboratory, Los Alamos, New Mexico 87545, USA}
\newcommand{\Lebedev}{Nuclear Physics Department, Lebedev Physical Institute, Leninsky Prospect 53, 119991 Moscow, Russia}
\newcommand{\LLL}{Lawrence Livermore National Laboratory, Livermore, California 94550, USA}
\newcommand{\LosAlamos}{Los Alamos National Laboratory, Los Alamos, New Mexico 87545, USA}
\newcommand{\Manchester}{School of Physics and Astronomy, University of Manchester, Oxford Road, Manchester M13 9PL, United Kingdom}
\newcommand{\MIT}{Lincoln Laboratory, Massachusetts Institute of Technology, Lexington, Massachusetts 02420, USA}
\newcommand{\Minnesota}{University of Minnesota, Minneapolis, Minnesota 55455, USA}
\newcommand{\Crookston}{Math, Science and Technology Department, University of Minnesota -- Crookston, Crookston, Minnesota 56716, USA}
\newcommand{\Duluth}{Department of Physics, University of Minnesota Duluth, Duluth, Minnesota 55812, USA}
\newcommand{\Ohio}{Center for Cosmology and Astro Particle Physics, Ohio State University, Columbus, Ohio 43210 USA}
\newcommand{\Otterbein}{Otterbein College, Westerville, Ohio 43081, USA}
\newcommand{\Oxford}{Subdepartment of Particle Physics, University of Oxford, Oxford OX1 3RH, United Kingdom}
\newcommand{\PennState}{Department of Physics, Pennsylvania State University, State College, Pennsylvania 16802, USA}
\newcommand{\PennU}{Department of Physics and Astronomy, University of Pennsylvania, Philadelphia, Pennsylvania 19104, USA}
\newcommand{\Pittsburgh}{Department of Physics and Astronomy, University of Pittsburgh, Pittsburgh, Pennsylvania 15260, USA}
\newcommand{\IHEP}{Institute for High Energy Physics, Protvino, Moscow Region RU-140284, Russia}
\newcommand{\Rochester}{Department of Physics and Astronomy, University of Rochester, New York 14627 USA}
\newcommand{\RoyalH}{Physics Department, Royal Holloway, University of London, Egham, Surrey, TW20 0EX, United Kingdom}
\newcommand{\Carolina}{Department of Physics and Astronomy, University of South Carolina, Columbia, South Carolina 29208, USA}
\newcommand{\SLAC}{Stanford Linear Accelerator Center, Stanford, California 94309, USA}
\newcommand{\Stanford}{Department of Physics, Stanford University, Stanford, California 94305, USA}
\newcommand{\StJohnFisher}{Physics Department, St. John Fisher College, Rochester, New York 14618 USA}
\newcommand{\Sussex}{Department of Physics and Astronomy, University of Sussex, Falmer, Brighton BN1 9QH, United Kingdom}
\newcommand{\TexasAM}{Physics Department, Texas A\&M University, College Station, Texas 77843, USA}
\newcommand{\Texas}{Department of Physics, University of Texas at Austin, 1 University Station C1600, Austin, Texas 78712, USA}
\newcommand{\TechX}{Tech-X Corporation, Boulder, Colorado 80303, USA}
\newcommand{\Tufts}{Physics Department, Tufts University, Medford, Massachusetts 02155, USA}
\newcommand{\UNICAMP}{Universidade Estadual de Campinas, IFGW-UNICAMP, CP 6165, 13083-970, Campinas, SP, Brazil}
\newcommand{\UFG}{Instituto de F\'{i}sica, Universidade Federal de Goi\'{a}s, CP 131, 74001-970, Goi\^{a}nia, GO, Brazil}
\newcommand{\USP}{Instituto de F\'{i}sica, Universidade de S\~{a}o Paulo,  CP 66318, 05315-970, S\~{a}o Paulo, SP, Brazil}
\newcommand{\Warsaw}{Department of Physics, University of Warsaw, Ho\.{z}a 69, PL-00-681 Warsaw, Poland}
\newcommand{\Washington}{Physics Department, Western Washington University, Bellingham, Washington 98225, USA}
\newcommand{\WandM}{Department of Physics, College of William \& Mary, Williamsburg, Virginia 23187, USA}
\newcommand{\Wisconsin}{Physics Department, University of Wisconsin, Madison, Wisconsin 53706, USA}
\newcommand{\deceased}{Deceased.}

\affiliation{\ANL}
\affiliation{\BNL}
\affiliation{\Caltech}
\affiliation{\Cambridge}
\affiliation{\UNICAMP}
\affiliation{\Cincinnati}
\affiliation{\FNAL}
\affiliation{\UFG}
\affiliation{\Harvard}
\affiliation{\HolyCross}
\affiliation{\Houston}
\affiliation{\IIT}
\affiliation{\Indiana}
\affiliation{\Iowa}
\affiliation{\UCL}
\affiliation{\Manchester}
\affiliation{\Minnesota}
\affiliation{\Duluth}
\affiliation{\Otterbein}
\affiliation{\Oxford}
\affiliation{\Pittsburgh}
\affiliation{\RAL}
\affiliation{\USP}
\affiliation{\Carolina}
\affiliation{\Stanford}
\affiliation{\Sussex}
\affiliation{\TexasAM}
\affiliation{\Texas}
\affiliation{\Tufts}
\affiliation{\Warsaw}
\affiliation{\WandM}

\author{P.~Adamson}
\affiliation{\FNAL}


\author{I.~Anghel}
\affiliation{\Iowa}
\affiliation{\ANL}



\author{A.~Aurisano}
\affiliation{\Cincinnati}









\author{G.~Barr}
\affiliation{\Oxford}









\author{M.~Bishai}
\affiliation{\BNL}

\author{A.~Blake}
\affiliation{\Cambridge}


\author{G.~J.~Bock}
\affiliation{\FNAL}


\author{D.~Bogert}
\affiliation{\FNAL}




\author{S.~V.~Cao}
\affiliation{\Texas}

\author{C.~M.~Castromonte}
\affiliation{\UFG}




% ...

\collaboration{The MINOS Collaboration}
\noaffiliation



\begin{abstract}
Kinematic distributions from an inclusive sample of $1.41\times 10^{6}$ charged-current $\nu_{\mu}$ interactions on iron, obtained using the MINOS Near Detector exposed to a wide-band beam with peak flux at 3\,GeV, are compared to a conventional treatment of neutrino scattering within a Fermi gas nucleus. 
Results are used to guide the selection of a subsample enriched in quasielastic $\numu$Fe interactions,
containing an estimated 123,000 quasielastic events of incident energies $1 < E_{\nu} <  8$\,GeV, with $\langle E_{\nu} \rangle = 2.79$\,GeV.  
Four additional subsamples representing topological and kinematic sideband regions to quasielastic scattering are also
selected for the purpose of evaluating backgrounds.  Comparisons using subsample distributions in four-momentum transfer $Q^2$ show the Monte Carlo 
model to be inadequate at low $Q^2$.  Its shortcomings are remedied via inclusion of a $Q^2$-dependent suppression function for baryon resonance production, developed from the data.   A chi-square fit of the resulting Monte Carlo simulation to the shape of the $Q^2$ distribution for the quasielastic-enriched sample is carried out with the axial-vector mass $M_{A}$ of the dipole axial-vector form factor of the neutron as a free parameter.  The effective $M_{A}$ which best describes the data is $1.23^{+0.13} _{-0.09} \mbox{(fit)} ^{+0.12} _{-0.15} \mbox{(syst.)}$\,GeV. 
\end{abstract}

\maketitle
\section{Introduction} \vspace{-9pt}
\label{sec:Intro}

Recent measurements of neutrino interactions in nuclei have challenged our understanding
of how neutrino-nucleon scattering is modified when the target nucleons are entangled within
a nuclear binding potential.
Cross section discrepancies relative to $\nu_{\mu}$N and $\anumu$N scattering on free
nucleons are particularly apparent for charged current (CC) neutrino-nucleus interactions
initiated by neutrinos in the energy range of $E_{\nu}$ 
from 0.5 to a few GeV~\cite{K2KMA:2006, Scibar:2007, 
MiniBooNEMA:2008,  NOMADMA:2009, MiniBooNEMA:2010, 
SciBooNEInclusive:2011, T2K-2013, Minerva-1, Minerva-2}.
Meanwhile,  the accuracy of neutrino interaction models is 
becoming increasingly important to the analysis of neutrino flavor
oscillation experiments, especially for CC interactions.
The detector configurations deployed in neutrino oscillation experiments have
given rise to new, high statistics neutrino scattering measurements on carbon and oxygen nuclei.
The results to date have made it clear that
models tuned primarily on light-liquid bubble-chamber data do not provide precise
descriptions of neutrino-nucleus interactions~\cite{Gallagher-ARNS, Huber-2013}.  

The present work seeks to shed light on CC
$\numu A$ scattering in the region $1 < E_{\nu} <  8$\,GeV.
For this purpose, an overview of inclusive CC scattering is established by comparing data
of selected event samples to the predictions of a conventional Monte Carlo (MC) treatment wherein 
neutrinos interact with the nucleons of a relativistic Fermi gas.   These samples are used
to guide an analysis of charged-current quasielastic (CCQE) scattering,
\begin{equation}\label{eq:quasielastic-reaction}
\nu_{\mu}+\mathrm{n} \rightarrow \mu^{-}+ \mathrm{p}  ~,
\end{equation}
\noindent
the fundamental semileptonic interaction that features prominently in many
neutrino oscillation measurements.

In contrast to nearly all previous works, the target neutrons of this study are bound
within the large iron ($A \simeq$ 56) nucleus~\cite{Kustom-1969}.
The neutrino energy spectrum analyzed here overlaps and extends
the $E_{\nu}$ region studied by 
K2K~\cite{K2KMA:2006,Scibar:2007}, MiniBooNE~\cite{MiniBooNEMA:2008, MiniBooNEMA:2010}, 
SciBooNE~\cite{SciBooNEInclusive:2011}, and T2K~\cite{T2K-2013}.  It extends to the beginning of the 
high energy region studied by NOMAD (with average $E_{\numu}$of 25.9\,GeV)~\cite{NOMADMA:2009};
it coincides very nearly with the $\nu_{\mu}$ 
spectrum investigated by MINER$\nu$A~\cite{Minerva-2}.    Thus, the observations pertaining to
CCQE interactions in $\numu$Fe collisions reported here are complementary to information gleaned
from CCQE scattering on $A \simeq 12, 16$ nuclei.

The neutrino interactions of this work were recorded in the MINOS Near Detector using an
exposure to the NuMI neutrino beam at Fermilab operated in its low-energy configuration.
Particular attention is devoted to event distributions in the (positive) square 
four-momentum transfer between the neutrino and the target nucleon,
$Q^2 =  - q^2 = - (k_\nu - k_\mu)^2 > 0$, where $k_{\mu} (k_{\nu})$ is the four momentum of the outgoing (incoming)
lepton.   High-statistics 
$Q^2$ distributions of selected CC samples are
compared to the predictions of conventional neutrino scattering phenomenology as
encoded into the MC event generator {\small NEUGEN3}~\cite{neugen} used by the MINOS experiment.   
The neutrino interaction model of {\small NEUGEN3} provides
an overall characterization of neutrino CC interactions at incident energies of a few GeV.  
The analysis makes use of fits to the shapes
of $Q^2$ distributions in the selected event samples.
Information about the event rate is not used to constrain model parameters
in order to avoid the sizable uncertainties associated with the  
absolute normalization of the neutrino flux.

The analysis uses a conventional neutrino-generator description of
final-state initiation and evolution.  In particular, CCQE signal events are taken to be 
quasielastic interactions on quasi-free neutrons prior to final-state interactions.   
This simplified, somewhat naive formalism allows CCQE scattering in iron to be parametrized using the axial-vector mass, $M_{A}$,
while avoiding the complexities inherently present in interactions on nuclei.    The downside is that the $M_{A}$ value determined
from the data is an effective parameter only indirectly related to the axial-vector form factor of the neutron.   However, since previous
work in this field was based on similar formalism, the approach taken here allows straightforward comparison of its determination
for $\numu$Fe scattering to previous $M_{A}$ results obtained with light target nuclei.
 
\section{Outline} 
\label{sec:outline}

The paper proceeds as follows:  Section\,\ref{sec:CCQE} summarizes 
the role of the axial-vector form factor and of the axial mass parameter, $M_{A}$,
in quasielastic scattering, and summarizes the recent experimental determinations 
of an effective $M_{A}$ for CCQE interactions in nuclear targets.
Sections~\ref{sec:Beam} and \ref{sec:Detector-Exposure} present the relevant
aspects of the NuMI neutrino beam, of the MINOS Near Detector, and of the data
exposure.  The neutrino interaction model 
used by the reference MC
is described in Sec.\,\ref{sec:Monte-Carlo}.    The main interaction categories invoked by the
model are quasielastic scattering,
CC baryon resonance production for which production of $\Delta(1232)$ states is predominant,
and CC deep inelastic scattering including low-multiplicity pion production.
Other relatively low-rate channels are also treated.

Upon isolation of a large CC inclusive data sample ($1.41\times 10^6$ events), the analysis commences
with comparisons of kinematic distributions to MC predictions (Sec.\,\ref{sec:CC-Inclusive-Sample}).
The MC categorizations serve to guide the extraction of four independent
subsamples from the inclusive sample, whose
events populate topological and kinematic sideband regions to 
CCQE scattering (Sec.\,\ref{sec:selected-subs}).

In subsamples containing abundant CC baryon resonance production, the MC
predicts event rates which exceed the data rates
for the region $ 0 < Q^{2} < 0.5$\,GeV$^{2}$.
The data of the sideband samples are used to develop a $Q^2$-dependent
suppression weight for baryon resonance production (Sec.\,\ref{subsec:Suppression}).
This weight is subsequently included as
a refinement to the MC model, thus modifying the predicted amount of baryon-resonance
background at low $Q^2$ in the CCQE enhanced sample.   
The latter sample, selected to be enriched in quasielastic events,
is presented in Sec.\,\ref{sec:CCQE-sample}.  

The effective axial-vector mass is determined by fitting the shape 
of the $Q^2$ distribution of the CCQE enhanced sample.
The data-fitting framework to do this is presented in Sec.~\ref{sec:Effective-MA}.   
Evaluation of the sources of systematic uncertainty 
for the $M_{A}$ determination is presented in 
Sec.\,\ref{subsec:Systematic-errors}.
Final results are given in Sec.\,\ref{sec:Results-Discuss} 
and implications are discussed.

\section{CC quasielastic scattering}
\label{sec:CCQE}

The CCQE differential cross section with respect to the squared four-momentum 
transfer between the leptonic and hadronic currents, $Q^{2}$, follows the general form
\begin{eqnarray}
\label{eq:CCQE-dif-x-section}
\begin{split}
\frac{d\sigma}{dQ^{2}} = &\frac{M_{n}^{2}G_{F}^{2}\cos^{2}(\theta_{c})}{8\pi E_{\nu}^{2}} \times \\
&\left\{A(Q^{2}) + B(Q^{2})\frac{(s-u)}{M_{n}^{2}}+C(Q^{2})\frac{(s-u)^{2}}{M_{n}^{4}} \right\}.
\end{split}
\end{eqnarray}

\vspace{2pt}

\noindent Here, $(s-u) = 4E_{\nu}M_{n} - Q^{2} - m_{\mu}^{2}$,  
where $M_{n}$ is the mass of the struck neutron and
$m_{\mu}$ is the mass of the final-state muon. 
The functional forms $A(Q^{2})$, $B(Q^{2})$, 
and $C(Q^{2})$ contain terms with various
combinations of the nucleon vector form factors 
and the nucleon axial-vector form factor $F_A(Q^2)$;  their explicit forms
are given in Ref.~\cite{LlewellynSmith:1972}.  
(For antineutrino CCQE scattering, $M_{n} \rightarrow M_{p}$ and 
the sign of $B(Q^{2})$ is reversed.)  
Additionally there are terms within $A(Q^{2})$ and $B(Q^{2})$
which contain the pseudoscalar form factor $F_P(Q^2)$.    
These however have a negligible effect 
in the present analysis and are ignored.
According to conventional phenomenology the vector form factors satisfy 
the conserved vector current (CVC) hypothesis~\cite{Feynman:CVC} 
and therefore are directly related to 
the Sachs electric and magnetic form factors~\cite{Sachs:1964}.   
The latter form factors have been well measured
by electron scattering experiments.   
The coupling strength of the axial-vector form factor
at zero four-momentum transfer, $F_{A}(Q^2=0)$,
is well known from neutron $\beta$-decay experiments. 
Consequently a full description of the differential 
cross section for CCQE scattering hinges upon determination 
of the axial-vector form factor, $F_{A}(Q^2)$.
The form factor's fall-off with increasing $Q^{2}$ 
is conventionally parametrized using the empirical dipole form
\begin{equation}
F_{A}(Q^{2}) = {F_{A}(0)}/{\left(1+ {Q^{2}}/{M_{A}^{2}}\right)^{2}}.
\end{equation} 
\noindent Thus the axial-vector form factor can be described with just one parameter,
the axial-vector mass, $M_{A}$.    The magnitude of $M_{A}$ determines 
the shape of the $Q^2$ momentum transfer spectrum
and sets the scale for the absolute CCQE cross section
$\sigma$($E_{\nu}$) (hence the total CCQE rate in an experiment) as well.

The value of $M_{A}$ can be extracted by measuring the $Q^{2}$ distribution for CCQE 
scattering events.    One decade ago, the world-average value 
for $M_{A}$ was (1.026\,$\pm$\,0.021)\,GeV \cite{Bernard:2002}; 
 this value was dominated by measurements obtained using
large-volume bubble chambers filled with liquid deuterium such as those operated at
the Argonne~\cite{ANL:1982} and Brookhaven~\cite{BNL:1990} National Laboratories.
More recent experiments use parametrizations of vector form factor measurements
obtained by electron scattering experiments,
a refinement that shifts $M_{A}$ to the lower value of
0.99\,GeV~\cite{Kuzmin-2008}.

In recent times, high-statistics experiments using tracking spectrometers have studied the CCQE
interaction using nuclear targets.   The K2K experiment reported an $M_{A}$ value
of (1.20$\,\pm$\,0.12)\,GeV using oxygen as the target nuclei~\cite{K2KMA:2006} and 
the MiniBooNE experiment measured (1.35\,$\pm$\,0.17)\,GeV 
using carbon as the target medium ~\cite{MiniBooNEMA:2010}.
On the other hand the NOMAD experiment, 
working in a distinctly higher $E_{\nu}$ range, obtained $M_{A} = (1.05 \pm 0.06$)\,GeV, a value
consistent with the bubble chamber results~\cite{NOMADMA:2009}.    A widely-held viewpoint
is that the apparent spread in $M_{A}$ values is driven 
by nuclear medium alterations of the CCQE free-nucleon cross section~\cite{Gallagher-ARNS, Morfin:2012kn}.

The MINER$\nu$A experiment has reported flux-averaged $d\sigma/dQ^{2}$ distributions 
for both neutrino and antineutrino quasielastic scattering on carbon~\cite{Minerva-1, Minerva-2}.   
The distributions span the energy range $1.5 <  E_{\nu} < 10$\,GeV and thereby bridge the
ranges examined by other recent experiments.   
Satisfactory fits are obtained for both data sets 
using a Relativistic Fermi Gas (RFG) nuclear model with $M_{A}$ = 0.99\,GeV.   
However, augmentation of the RFG model 
with the transverse enhancement model (TEM)~\cite{ref:Bodek-Budd-Christy} improves the description.   
The TEM involves a distortion to the magnetic form factors for bound nucleons, 
a phenomenological result extracted from electron-carbon scattering data 
and applied directly to the same magnetic form factor in the neutrino case.

There are other phenomenological models 
which deduce the effects of the nuclear medium on CCQE scattering 
based on knowledge of electron-nucleus scattering.   
These models build upon RFG and include the final-state suppression resulting 
from Pauli exclusion of the reaction proton from occupied levels of the target nucleus.   
The effect of Pauli blocking on CCQE is significant; in a large nucleus such as iron 
the suppression extends from 0.0 to $ \sim 0.3$\,GeV$^{2}$ in $Q^{2}$.

Current models include contributions due to multi-nucleon effects 
such as nucleon-nucleon correlations and
two-particle two-hole (2p2h) processes~\cite{Nieves:2004wx, Martini:2009uj, Martini:2010ex, 
Martini:2011wp, Amaro:2010sd, Megias:2013, Nieves:2011yp, Gran:2013kda}.
These additional processes can initiate scattering events which, 
in many experiments, would appear to be CCQE-like and would distort the CCQE cross section.    
In one approach, the differential and total cross sections for CCQE are calculated as
the squared sum of all microscopic interaction amplitudes devoid of pion emission, 
including reaction~\eqref{eq:quasielastic-reaction}~\cite{Martini:2009uj, Nieves:2011pp, Gran:2013kda}.   
Another approach is to use scaling arguments 
to estimate component contributions in electron scattering data and then to apply them 
to neutrino processes~\cite{Amaro:2010sd, Megias:2013}.   
Among recent works, Ref.~\cite{Gran:2013kda} obtains a description of
MINER$\nu$A data comparable to the models presented with those measurements, 
while also describing the MiniBooNE measurement~\cite{Nieves:2011yp}.   

In a recent measurement reported by MINER$\nu$A, a conventional Fermi gas treatment of nuclei is found to give
a poor description of the evolution of event rates with target $A$ (from C to Fe to Pb), for CC scattering samples having
sizable quasielastic contributions~\cite{Minerva-3}.
This observation suggests that nuclear medium effects may become more pronounced in
neutrino CCQE scattering when relatively large nuclei are used.   Given that most of the
phenomenological approaches described above are applicable to larger nuclei, new measurements
of CCQE scattering from a large-$A$ nucleus such as iron are of keen interest.

\section{N$\mathrm{u}$MI neutrino beam}
\label{sec:Beam}

The neutrino beam used in this measurement is produced by the NuMI facility at Fermilab~\cite{numi}. 
Protons with energy of 120\,GeV are extracted from the Main Injector accelerator in an 8\,$\mu$s spill 
every 2.2\,s and directed 
onto a graphite target of length corresponding to 2.0 proton interaction lengths.
The downstream end of the target was inserted 25\,cm into
the neck of the first (most upstream) of two focusing horns consisting of 
pulsed air-core toroidal magnets operated with a peak current of 185 kA.
Positively charged pions and kaons produced in the target 
are focused towards the beam axis by the magnetic horns, and are 
directed into a 675\,m long evacuated decay pipe. 
The neutrinos are produced by the subsequent decays of the mesons,
 as well as by decays of some of the daughter muons. 
The decay region is terminated by a hadron absorber.   
Residual muons are ranged out in the 240\,m of rock 
between the absorber and the Near Detector.    The Near Detector is located 1.04\,km 
downstream from the target in a cavern 100\,m underground. 

To predict the neutrino flux and consequent event rate spectrum, 
the simulation package {\small FLUKA05}~\cite{fluka05} is used to calculate the production 
of secondary hadrons created by the collision of primary protons with the graphite target.   
The transport of these hadrons and of their decay products (primarily neutrinos, 
muons, pions, and kaons) along the
NuMI beamline is then calculated using {\small GEANT3}~\cite{geant3}.
Interactions of the neutrinos striking the Near Detector are simulated 
using the {\small NEUGEN3}  neutrino event generator.  

 Refinements to the {\it ab initio} simulation 
 of the beam flux at the detector are subsequently made 
 using a fitting procedure in which the energy spectra of CC interactions 
 observed in the detector are compared to predictions of the MC beam simulation.
 For this purpose, runs of short duration were taken 
 in which the primary target was situated at positions displaced longitudinally 
 from the nominal; the horn currents were also varied.
 With each distinct configuration of target location and horn currents, 
 data are obtained for which the transverse and longitudinal momentum spectra 
 of the hadrons focused by the horns, are modified.
 Consequently the energy spectra for neutrino CC interactions 
 at the detector are different for each run.    This enables the simulated descriptions of
  transverse and longitudinal momentum distributions 
  of  produced $\pi$ and $K$ mesons to be adjusted so as to obtain the best agreement with 
  the neutrino event rate spectrum of each run~\cite{prd_1e20}.   
  Hereafter, the MC calculation with the modifications 
  described above is referred to as the {\it flux-tuned MC}.
    
  An independent estimation of the neutrino flux for $E_{\nu} > 3.0$\,GeV was carried out using
  a CC subsample characterized by low hadronic energy, to determine the flux shape~\cite{ref:CCinclusive}.
  The flux obtained is consistent throughout its $E_{\nu}$ range with the flux tune used by the present work.

\begin{figure}
\begin{center}
\includegraphics[width=8.6cm]{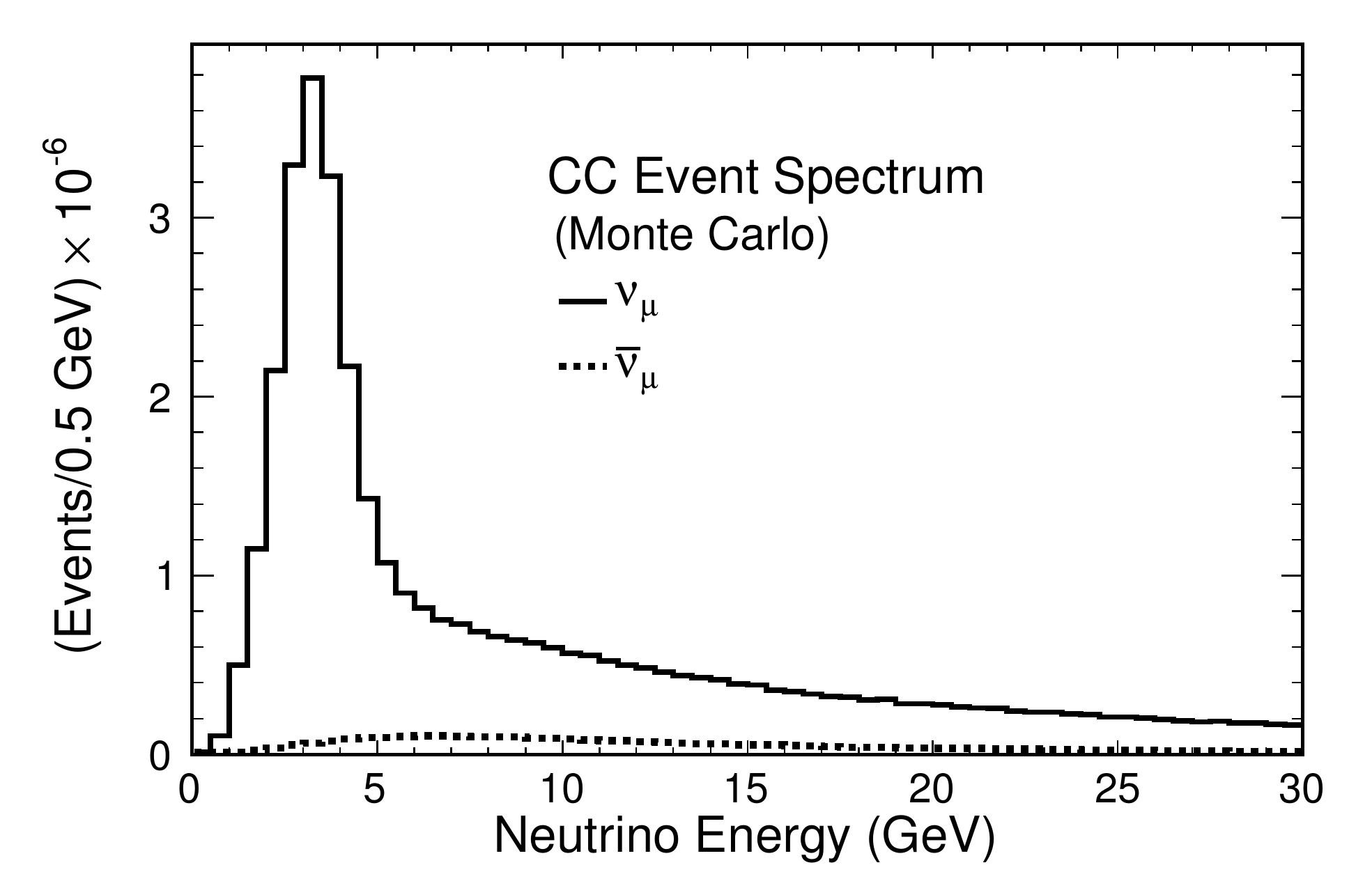}
\caption{Event rate spectra calculated for $\nu_{\mu}$ (solid line) and 
$\anumu$ (dashed line) CC interactions in the Near Detector.   
The quasielastic events and other low multiplicity CC interactions 
selected by this analysis arise predominantly from the $E_{\nu}$
region of 1.5 to 6.0 GeV.
\label{fig:ndspec}}
\end{center}
\end{figure}

 The neutrino CC event energy spectrum for the entire data exposure, calculated using
 the flux-tuning procedure described above, is displayed in Fig.~\ref{fig:ndspec}.  Also
 shown is the small contribution (dashed line) arising from antineutrinos in the beam. 
 For CC events in the Near Detector, the relative rates among neutrino flavors is estimated to be
 91.7\% $\nu_\mu$, 7.0\% $\anumu$, and 1.3\% $\nu_e$ + $\overline{\nu}_e$.
 
\section{Detector and data exposure}
\label{sec:Detector-Exposure}
\subsection{MINOS Near Detector}
\label{subsec:detector}

The Near Detector is a coarse-grained, 
magnetized tracking calorimeter composed of planes 
 of iron and plastic scintillator~\cite{ref:CCinclusive, minosNIM:2008}.  The    
 bulk of its 980 metric ton total mass resides in 282 vertically-mounted steel plates. 
 The upstream portion consisting of 120 planes comprises the detector's calorimeter section, 
 while the remaining 162 planes deployed downstream
 serve as the detector's muon spectrometer.   
 Each steel plate is 2.54 cm thick and corresponds 
 to 0.15 nuclear absorption lengths and 1.4 radiation lengths.
The scintillator planes are made of strips, 1 cm thick and 4.1 cm wide 
(1.1 Moli\`{e}re radii), oriented at $\pm$45$^{\circ}$ with respect to 
the vertical and alternating $\pm$90$^{\circ}$ in successive planes.   
The strips are read out with wavelength shifting fibers 
connected to multi-anode photo-multiplier tubes (PMTs). 
The 120 planes of the calorimeter span 
a distance of 7.2\,m along the beam direction, and the 
162 planes of the spectrometer extend 
the tracking volume by an additional 9.7\,m.
(The pitch of the detector is 5.97\,cm; 
it encompasses steel, scintillator module, and air gap.)
 In the muon spectrometer section, every fifth plane 
 is instrumented with the scintillator but the intervening planes
are bare steel with air gaps.

\begin{figure}
\begin{center}
 \includegraphics[width=8.4cm]{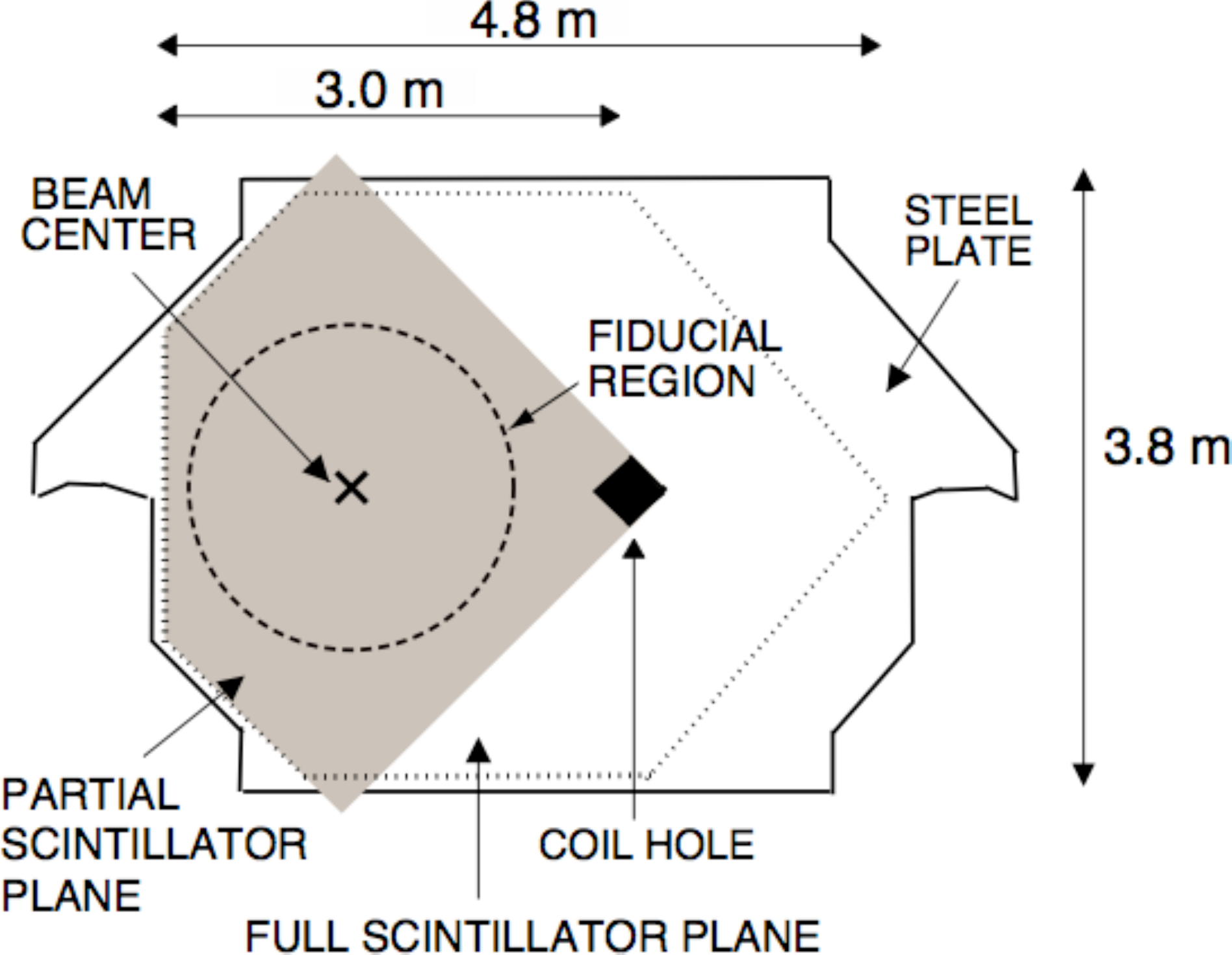}
\caption{Upstream transverse-face view of the MINOS Near Detector.  
The neutrino beam is centered between the axial-magnet coil hole 
and the left side of the stack of steel planes. 
The vertex fiducial volume is coaxial with the beam spot and begins
at a longitudinal depth of 1.0\,m within the calorimeter section.   
Every fifth steel plane is instrumented with a full scintillator plane 
(denoted by the dotted hexagonal border) while each of the four intervening planes 
has scintillator coverage as shown by the shaded region.  
The muon spectrometer section lies immediately downstream.
\label{fig:near-det-trans}}
\end{center}
\end{figure}

The steel planes are magnetized with a toroidal field 
of average strength 1.3\,T, arranged to focus negative muons.   
The magnetic field enables the charges of final state muons to be identified 
and provides a measurement of their momenta based upon
track curvature.    For muons that stop within the detector, 
the stopping distance provides an alternate, more accurate measure
of the track momentum at the interaction vertex.    
In this work, events with exiting muons are included in the kinematic sideband samples,
however muons of candidate quasielastic events are required to be
negatively-charged stopping tracks.

In the calorimeter section, every fifth plane is instrumented with a scintillator layer, while  
each of the four intervening planes has partial scintillator coverage
over the area transverse to the beam.   This is because, 
as shown in Fig.~\ref{fig:near-det-trans}, the neutrino beam 
is centered between the axial hole that carries the magnet coil 
and the left side of the planes, and so the scintillator 
only needs to cover this area~\cite{minosNIM:2008}.   

The relative locations of the event vertex fiducial region, 
the neutrino beam spot, the steel planes with the two types of scintillator coverage, 
and the magnet coil hole along the axis of the steel stack, are shown in Fig.~\ref{fig:near-det-trans}.

The neutrino interactions accepted for analysis occur in the forward part of the calorimeter section, 
in a fiducial volume defined as between 1 to 5 meters 
from the upstream end of the detector and within a 1 meter radius about the beam axis. 
The calorimeter section records the energy deposited by neutrino-induced hadronic showers.   
For the sub-GeV hadronic showers of interest to this analysis, the resolution for calorimetric measurement
of hadronic shower energy is approximately 80\%~\cite{minosNIM:2008, Dorman-Ehad-2008}.
The downstream spectrometer section provides
the curvature and range information required for reconstructing the momenta of muons 
from CC interactions in the calorimeter.   For a 3.0\,GeV muon, the energy resolution 
is 4.6\% for measurement by range.  Measurement by curvature has poorer resolution (11\%)
and is not used for the muon track reconstruction of candidate quasielastic events.  

 With measurement of track momentum 
from range, uncertainties arise from the detector mass,
from approximations to the detector geometry 
used by the reconstruction software, and from the model of energy loss.
These effects combine to give a 2\% systematic error for range-based momentum.

\subsection{Exposure, signal readout, calorimetric response}
\label{subsec:exposure}

The data of this analysis are from an exposure totaling $1.26 \times 10^{20}$ POT,  taken in
the first year of NuMI operation during 2005-2006.   The average proton intensity 
was $2.2 \times 10^{13}$ POT per accelerator
spill of 8\,$\mu$s duration.   At this intensity, an average of eight neutrino interactions 
occur in the calorimeter region during each spill.  For Near Detector $\numu$ CC samples at this exposure
the systematic errors  of the measurement dominate the statistical errors (see Sec.~\ref{subsec:Systematic-errors}).

To distinguish individual neutrino events in the detector from one another,
both timing and spatial information are used.  The readout electronics 
operate with essentially zero dead time.   The PMT signals are 
continuously digitized throughout the spill in contiguous 18.8\,ns intervals 
corresponding to the 53 MHz RF of the Main Injector.

Details concerning the calibrations required to convert raw PMT signals into deposited energy 
are given in Ref.~\cite{caldet}. 
The detector response to charged-particle traversal
was measured by MINOS using a scaled-down 12\,ton calorimeter having the same
composition and granularity as the MINOS detectors.    This replicate detector
was exposed to beams of protons, pions, muons, and electrons in the 
momentum range 0.2 to 10\,GeV, in a dedicated experiment at the CERN-PS~\cite{caldet2}.

\section{Reference Monte Carlo}
\label{sec:Monte-Carlo}

The MINOS neutrino event generator {\small NEUGEN3} provides descriptions of all
the neutrino scattering processes that contribute to the event rate 
in the $E_{\nu}$ regime of this study.  These include quasielastic scattering, 
baryon resonance production, low-multiplicity pion production, 
deeply-inelastic scattering (DIS), and coherent pion production.
The {\small NEUGEN3} models for these processes are nearly identical to
those of the {\sc genie} (version 2.6.0) neutrino event generator~\cite{genie}.    Similar
cross section categorizations are employed by other neutrino event generators currently in use
such as {\sc nuance}~\cite{ref:NUANCE}, {\sc neut}~\cite{ref:NEUT}, and {\small NuWro}~\cite{ref:NuWro}.

For quasielastic scattering, {\small NEUGEN3} uses the {\small BBBA05} parametrization~\cite{BBBA:2005} 
of the nucleon vector form factors and the empirical dipole form for the axial-vector form factor computed
with $F_{A}(0) = -1.267$ and with a nominal value for $M_{A}$ of 0.99\,GeV.
A relativistic Fermi gas  model of the nucleus includes the effects of Fermi motion and Pauli blocking.  
The RFG model is augmented with inclusion of a 
high-momentum tail to the distribution of nucleon momentum
as proposed by Bodek and Ritchie~\cite{bodekritchie}. 
This phenomenological augmentation allows a small number 
of MC events to exhibit kinematics which would not normally ensue
with an RFG model; the occurrence of such events 
is predicted by spectral function models~\cite{ref:Benhar-1994}.
In the generation of $\nu$Fe interactions by {\small NEUGEN3}, 
Pauli blocking is implemented as a rejection imposed upon generated
quasielastic and elastic interactions whose recoil protons 
(or neutrons) are below 251 MeV/c (below 263 MeV/c).
For generated events that survive the Pauli blocking step, 
the final-state hadrons are then propagated through the nuclear
medium and probabilities are assigned to the possible rescatterings 
according to an intranuclear cascade algorithm {\sc intranuke}~\cite{inuke_data}.
Via this particle cascade model, the detailed effects 
of pion and nucleon rescattering processes such as elastic and inelastic scattering, absorption,
and charge exchange scattering are accounted for in simulations carried out by the reference MC.

In generation of neutrino-induced baryon resonance production decaying
into the two-body final states (lepton + $\Delta$/N$^{*}$), {\small NEUGEN3} 
uses the phenomenological treatment of Rein and Sehgal~\cite{ReinSehgal:1981}.  
This formalism takes into account the
production of 18 different baryonic states in exclusive-channel reactions;  
the largest cross sections are those involving the charge states of
the $\Delta(1232)$ resonance. 
For the $\Delta(1232)$ and for other baryon resonances as well, 
the axial-vector form factor is taken to be the empirical
dipole form but with a mass value of $M_A^{RES} = 1.12$\,GeV~\cite{Kuzmin:2006dh}.   
For the present analysis (as with other MINOS studies),
the {\small NEUGEN3} generator does not impose Pauli blocking 
upon baryon resonance production.
In the decay of the various resonance states, 
the emission of the daughter particles is assumed to proceed isotropically in the
rest frame of the parent particle.

For its description of deep inelastic scattering,  {\small NEUGEN3} 
uses the formalism of Bodek and Yang~\cite{BodekYang:2004}, 
including the extension of the formalism that improves the modeling 
of the transition region from resonance to DIS interactions~\cite{neugen-dis:2007}.
A survey of neutrino interaction data from previous experiments was used to determine 
the appropriate hadronic mass spectrum ($W$ spectrum) to use with events in this transition
region.   For events having hadronic mass
between $1.7 < W < 2.0$\,GeV, a good match to data distributions for $d\sigma/dW$ is
achieved by implementing a linear evolution 
from the Rein-Sehgal exclusive channel treatment to the Bodek-Yang DIS model.
For the production of multiparticle hadronic systems 
as occurs with DIS events, two different approaches are employed.
For production of relatively low hadronic masses, final-state particle 
multiplicities are simulated according to a modified form of 
KNO scaling~\cite{KNO:1972}.   For higher invariant masses, 
e.g. $W$ $>$ 2.3\,GeV, the KNO hadronic shower model is evolved 
into a {\sc pythia/jetset} description~\cite{Sjostrand:2006}.  

The production of single pions via CC
coherent scattering on iron is a background for the present study, 
for events in which the final state pion goes undetected. 
{\small NEUGEN3} simulates this process using
an implementation of the PCAC-motivated coherent scattering model 
of Rein and Sehgal~\cite{ReinSehgal:1983}.
The cross section for this process is known to be small, 
and in fact its contribution as a background into the 
the lowest $Q^2$ bins for selected CCQE events is estimated 
by {\small NEUGEN3} to be 1\%.  The systematic error
arising from the particular coherent scattering implementation 
is negligible compared to other errors in this analysis.

The reference MC uses a materials assay of the MINOS Near Detector 
to determine its nuclear composition.   According to the MC, approximately 5\% of the neutrino interactions 
 recorded in the detector occur not on iron, but on the plastic scintillator and its aluminum skin.  
This 5\% contribution to the event rate is included in the MC simulations with appropriate modifications 
made to the nuclear Fermi gas model for the carbon, hydrogen, and aluminum nuclei, and to the treatment 
of intranuclear rescattering in these lighter, smaller nuclei.

\section{CC $\numu$ Inclusive Event Sample}
\label{sec:CC-Inclusive-Sample}

\subsection{Selection of the sample}
\label{subsec:select-cc-inclusive}

The foundational sample for the analysis is an inclusive sample of $\nu_{\mu}$ CC events selected from the data;
the same selections are applied to a realistic Monte Carlo simulation of the experiment.
The selection criteria used are mostly those used previously for the MINOS measurement of 
$\nu_{\mu}$ disappearance oscillations and are described in detail 
in~\cite{prl_3e20, rustem_thesis}.  In brief, the presence or absence
of a muon track in each event is ascertained using 
a multivariate likelihood discriminant.   The discriminant assigns a probability for the
muon hypothesis based upon four measured variables, 
namely the average pulse-height per plane along the track,
the transverse energy deposition along the track, the fluctuation of the energy deposits
strip by strip along the track, and the length of the track.    
A track is required to traverse six or more scintillator planes, giving a muon energy threshold of 300 MeV.   
The sample includes events with exiting as well as stopping muon tracks.

To the above selections, the analysis adds additional data quality requirements:  ({\it i\,})  A timing isolation cut is imposed; 
events in the calorimeter section that are concurrent within 70\,ns with another event are excluded.
This criterion eliminates instances of event pileup which occasionally lead to erroneous reconstructions.~({\it ii\,}) Events for which the
muon track either ends on the far side of the axial hole which carries the energizing coil 
(transverse locations to the right of the coil hole in Fig.~\ref{fig:near-det-trans}) or else stops within 45\,cm of its center,
are rejected.   

The efficiency with which CC events are selected is found to be 87\% in the MC simulation.   The detection
efficiency remains nearly constant with increasing muon angle with respect to the beam up to 35$^{o}$
and falls off rapidly at larger angles.   The events retained include 92\% of genuine CCQE events and 85\%
of two-body CC final states  $\mu^{-} + \Delta$/N$^{*}$.

\subsection{Kinematic variables; muon angular resolution}
\label{subsec:Muon-Ang-Res}

The recoiling hadronic systems of CC events often give rise to scintillator ``hits"  
that are clearly associated with the events but are not ionizations due to the muon tracks. 
The total summed pulse height from the hadronic shower hits in an event is used 
to estimate the system energy, hereafter designated as $E_{had}$. 

Prerequisites for this estimation are parametrizations of the detector response to energies of single hadrons and photons.  
Such parametrizations have been developed by MINOS;  they are based on simulations that have been
cross-checked against calibration data obtained from test beam exposures 
of a replicate detector to protons, pions, and electrons~\cite{minosNIM:2008}. 
The mapping of pulse height to $E_{had}$ is completed using modeling of hadronic showers.
The detector-response parametrizations are used in conjunction with estimations of 
the particle content of CC-induced hadronic systems, e.g. the particle types, multiplicities, and energies.
For CC events having $200 < E_{had}  < 250$ MeV, a region of particular interest to the analysis,
the mean multiplicities per event (according to the reference MC)
are 1.9 protons, 1.3 neutrons, 0.4 $\pi^{\pm}$, and 0.1 $\pi^{0}$ mesons.

For simulated CCQE events in this $E_{had}$ range,
the mean of reconstructed $E_{had}$ values falls within 6\% of the mean for MC true values.   

\begin{figure}
\begin{center}
\includegraphics[width=8.6cm]{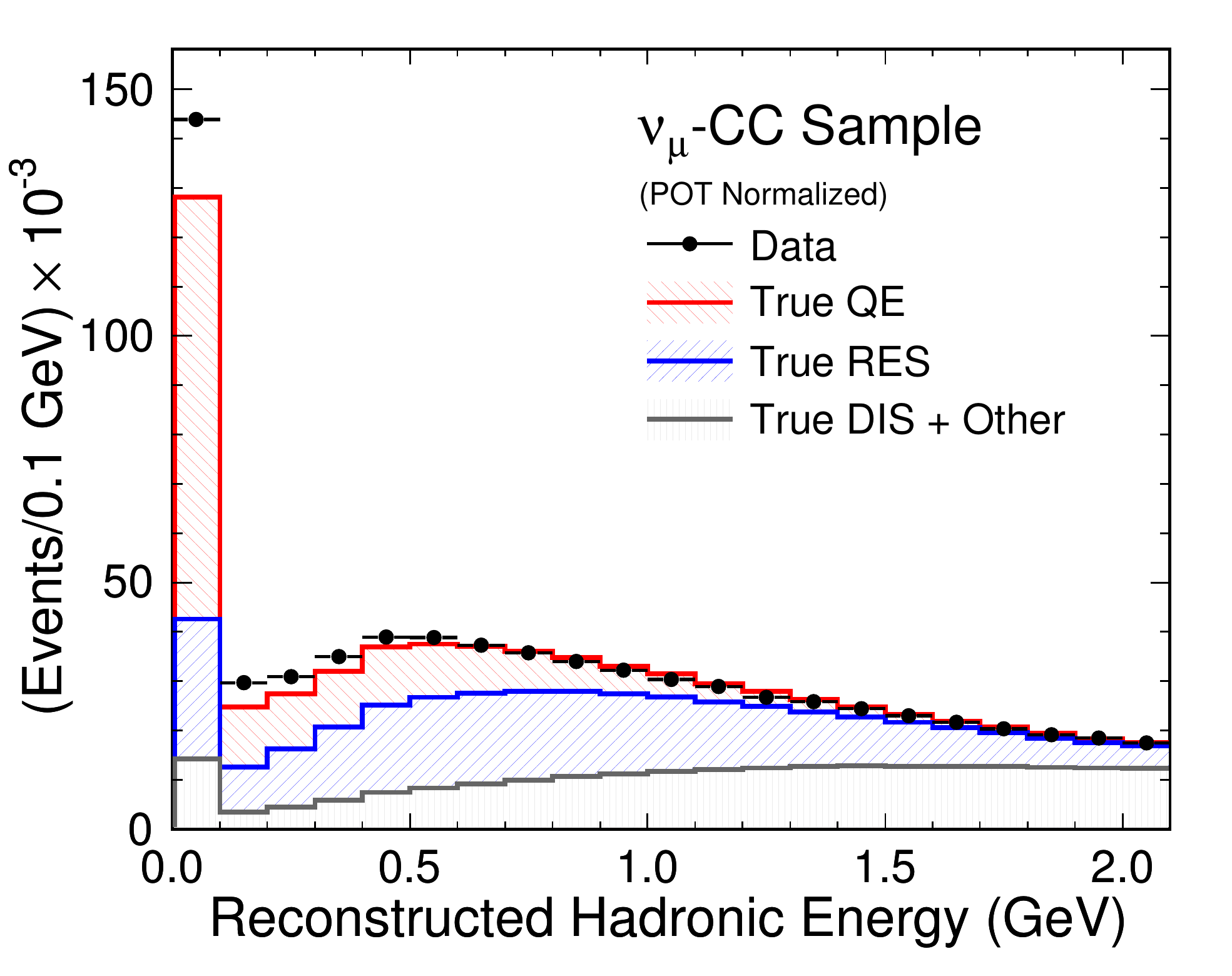}
\caption{Distribution of final-state hadronic energy, $E_{had}$, recoiling from the muon track in events of the inclusive CC samples
of data (solid circles) and the flux-tuned MC (histograms).    Subsamples of CCQE and CC baryon resonance channels
as estimated by the MC are shown by the two elevated, hatched histograms (distributions are stacked).  
The remaining subsample (lowest, shaded) arises  
from CC non-resonant pion production and low-multiplicity DIS reactions.  
 \label{fig:ccSample_eshw}}
 \end{center}
\end{figure}

The distribution of reconstructed energies of the final-state hadronic systems, 
$E_{had}$, in events of the CC inclusive sample is shown in 
Fig.~\ref{fig:ccSample_eshw}.   The data events (solid circles) 
are displayed together with the predictions from the 
flux-tuned MC for the same total exposure.   
 For $E_{had}$ bins above 500 MeV, the MC prediction
 agrees with the hadronic energy distribution.   For the lower
 energy bins however, where CCQE is the dominant interaction, 
 there is a relative excess of data;  the MC underestimates the data rate by $\sim 11\%$.

The extent to which the reference MC simulation describes the CC inclusive sample is of general interest.
Additional comparisons are afforded by the following kinematic estimators:
\begin{equation}\label{eq:E-nu-est}
E_{\nu}^{(est)} =    E_\mu   +  E_{had}~,
\end{equation}
\noindent and
\begin{equation}\label{eq:Q2-without-QE}
Q^{2} = 2E_{\nu}^{(est)}(E_{\mu}-p_{\mu}\cos\theta_{\mu})-m_{\mu}^{2}~.
\end{equation}

From Eq.~\eqref{eq:Q2-without-QE} it can be seen that reconstruction of
the muon angle, $\theta_{\mu}$, is important for the calculation of $Q^{2}$.
In MINOS, the resolution, $\sigma(\theta_{\mu})$, ranges between 16\,mrad and 52\,mrad for muons of the 
highest momenta (long tracks) and lowest momenta (short tracks), respectively.
The angular resolution in the MC was compared to the data using two different methods.
 In one method the reconstructed angles of upstream versus downstream segments 
were compared at track midpoints, separately for muons from the data and the MC.    
In the other method, angles of track segments reconstructed 
in the Near Detector were compared with the reconstructed angle 
of the same tracks as they exit the upstream MINER$\nu$A detector~\cite{minerva-nim}.   
Both methods showed the data to have better muon angular resolutions
than were represented in the MC.   The discrepancy was largest for short tracks 
(p$_{\mu} <$ 2\,GeV/c, $\overline{\sigma} =52$\,mrad), with reduction (in quadrature) of $\sim$14\,mrad of smearing 
from the MC being required to match the data.   However it diminished steadily with increasing track length, 
becoming indiscernible for tracks longer than 130 sampling planes (p$_{\mu} >$ 5\,GeV/c).   
This effect was shown to be unrelated to uncertainties with detector alignment;  rather, it is attributed to 
cumulative errors in the MC model of detector response.

The MC-vs-data angular resolution discrepancy gives rise to a mild flattening of MC $Q^2$ distributions whose form is determined
as follows:  A randomized smearing of reconstructed angles is applied to muon tracks of data to obtain a sample having the 
resolution of the MC.   The ratio of original to smeared data is constructed in bins of $Q^2$; the ratio exhibits a regular
dependence which is well-described using a polynomial function.    The MC (bin-by-bin in $Q^2$) is then divided by values of the ratio
function to obtain an MC distribution that would ensue if its resolution was identical to that of the data.   Thus the ratio function serves as a 
correction weight which, hereafter, is applied to individual MC events according to their $Q^2$ values~\cite{QE-Position-Paper}. 

Ratio functions are determined separately for the CC inclusive sample, for the sideband subsamples, and for the 
CCQE enhanced sample;  however there are only small differences among these functions.
For all samples,  the correction to the MC $Q^2$ distribution amounts 
to $3\%$ as $Q^{2}$ approaches 0.0 and $< 2\%$ for all higher values.  
The uncertainties in MC-vs-data resolution differences 
per bin of track length imply a range-of-variation allowed to the 
correction weight.   The one-sigma error band calculated for the weight is used
to assign a systematic error to this correction.

The resolution in $Q^2$ (or $Q^2_{QE}$ of Eq.~\eqref{eq:Q2-QE}), is as follows:  
For $Q^2 (Q^2_{QE})$ below 0.05 GeV$^2$, the resolution is 0.03 (0.02) GeV$^2$.
At larger $Q^2 (Q^2_{QE})$ the resolution increases to 0.08 (0.07) GeV$^2$ 
at 0.25 GeV$^2$ and 0.13 (0.11) GeV$^2$ at 0.45 GeV$^2$.
For $Q^2 >$ 0.5\,GeV$^2$, the fractional resolution, 
$(Q^{2}_{reco} - Q^{2}_{true})/ Q^{2}_{true}$, is 
constant at 28\% (25\%) of $Q^2 (Q^2_{QE})$~\cite{dorman-thesis}.

\subsection{Kinematic distributions: Data versus the MC}
\label{subsec:cc-incl-distributions}

\begin{figure}
\begin{center}
\includegraphics[width=8.3cm]{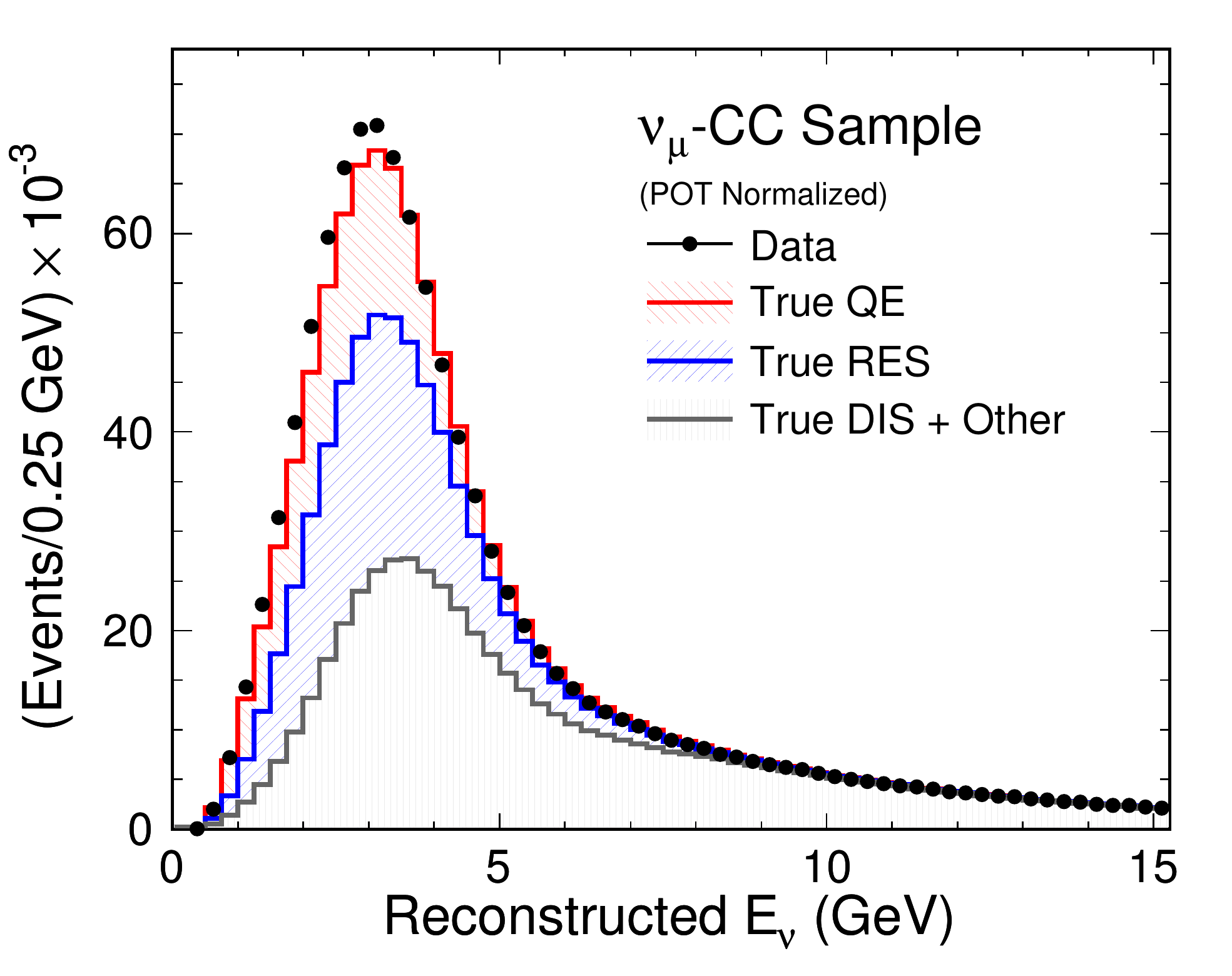}
\includegraphics[width=8.3cm]{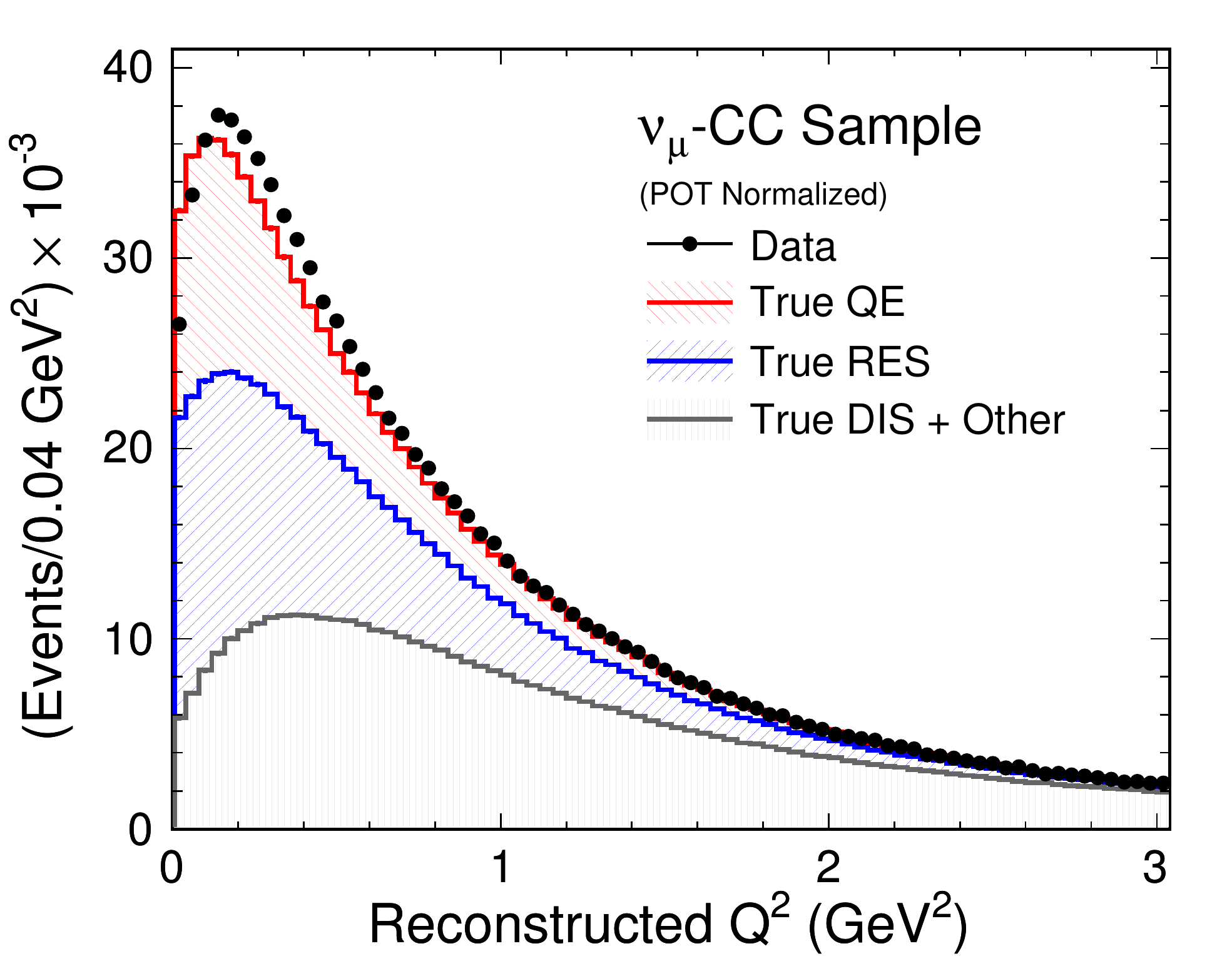}
{\caption{Distributions of the CC inclusive sample in data (solid circles) 
and from the MC (histograms),  for 
reconstructed neutrino energy (top) and for $Q^{2}$ (bottom).   
The component histograms (hatched, stacked) show MC predictions
for contributions arising from quasielastic,  baryon resonance production, and deep-inelastic
scattering channels.   The MC 
underestimates the rising edge of the neutrino energy distribution in data (top), 
and exhibits deviations from the shape of the data
$Q^{2}$ distribution (bottom).
\label{fig.ccSampleEnuAndQ2}}}
\end{center}
\end{figure}

Figure~\ref{fig.ccSampleEnuAndQ2} compares the CC inclusive data to the MC prediction
for event distributions in reconstructed $E_{\nu}$ and $Q^{2}$ (upper, lower plots respectively).  
The flux-tuned MC prediction is normalized using the total protons-on-target for the data exposure (POT normalization).
The relative contributions from quasielastic scattering and from the two other
dominant interaction categories
are shown by the component (hatched) histograms. 
The MC (histogram upper boundary) is seen to provide first-order characterizations 
of the data distributions (solid circles).

\begin{figure}
\begin{center}
\includegraphics[width=8.6cm]{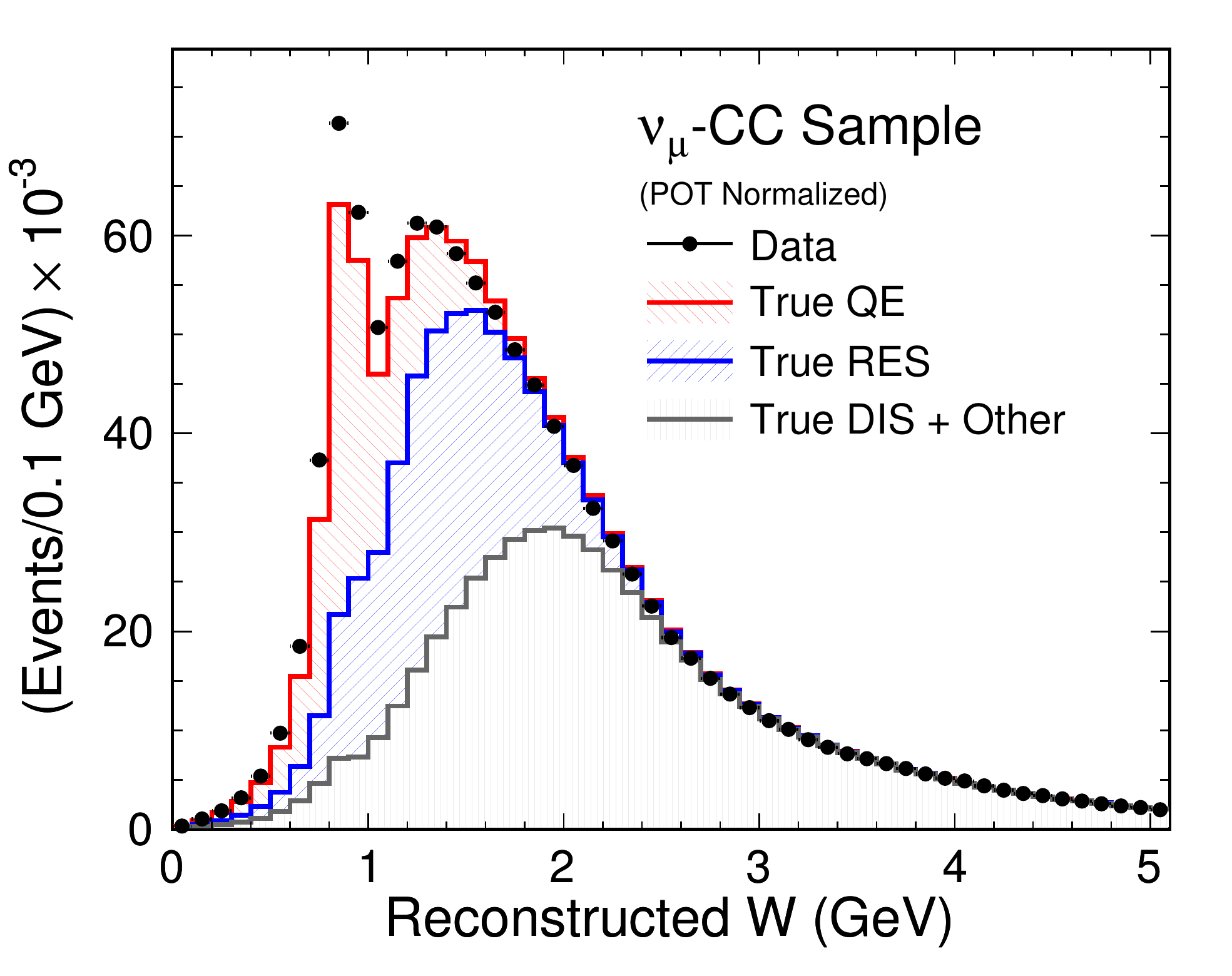}
{\caption{Distribution of reconstructed hadronic mass $W$ for events of the 
 CC inclusive sample, compared to MC predictions for the main interaction categories
 (stacked histograms, normalized to the data exposure).   The data exhibits an
 apparent excess relative to the MC in the region of the quasielastic peak;
 this feature is strongly correlated with the apparent data excess
 at $E_{had} \simeq 0.0$ GeV seen in Fig.~\ref{fig:ccSample_eshw}.
  \label{fig.CC-Inc-W-distribution}
}}
\end{center}
\end{figure}

A modest but useful degree of separation among the quasielastic, baryon resonance, 
and deep-inelastics scattering categories 
is provided by the final-state hadronic mass, $W$, reconstructed event-by-event in this 
analysis using the relation
\begin{equation}\label{eq:hadronic-W}
W^2 = M^2_{n} + (2\;M_n\;E_{had})-Q^2~.
\end{equation}
\noindent
Figure~\ref{fig.CC-Inc-W-distribution} shows the distribution of hadronic system invariant mass 
for the CC inclusive sample.   MC predictions for the three major interaction categories
are shown as stacked histograms; the predicted event rates are normalized to the data exposure.   
The fractional resolution is linearly proportional to $W$ through the region 0.6 to 2.0\,GeV,
improving gradually with increasing $W$ from 32\% to 20\%.
Clearly discernible is the quasielastic peak at $W$ values near the nucleon mass.   The peak 
is comprised of low-$Q^2$ events with $E_{had}$ approaching zero; 
event reconstruction smearing extends the data and MC distributions to values below $M_{n}$.    

The top plot in Fig.~\ref{fig.ccSampleEnuAndQ2} shows the residual 
data-vs-MC disagreement after the flux-tuning procedure.  
The flux tuning uses beam optics and hadron production parameters 
to obtain the apparent agreement with the total event rate at high energy ($E_{\nu}$ above 6.0 GeV).   
The lack of agreement around the spectral peak 
is not readily attributable to either the flux or cross section models, 
since the tuning parameters will tend to compensate for shortcomings with either one.    
Note that CC DIS is the dominant process at high energy, 
and that data-vs-MC agreement correlates with the DIS event rate 
in all the distributions of Figs.~\ref{fig:ccSample_eshw},  \ref{fig.ccSampleEnuAndQ2}, and \ref{fig.CC-Inc-W-distribution}.   
On the other hand, the apparent data excess relative to the MC around the spectral peak, 
where CCQE and baryon-resonance production account for a large fraction of the event rate, 
correlates with apparent excesses in related regions of the other figures.

Despite the flux-tuning procedure, the data in Fig.~\ref{fig.ccSampleEnuAndQ2}\,(bottom)  exhibit 
a sharper falloff as $Q^2$ approaches zero than is predicted by the MC.  
This latter feature cannot be explained by uncertainties in the neutrino flux, or by uncertainties
in the energy-dependence of exclusive-channel cross sections, $\sigma(E)_{QE,RES}$.
Rather, such an effect is more naturally related to nuclear medium effects and/or form factor behavior,
which is the physics targeted by the present analysis.

\section{Selected CC Subsamples}
\label{sec:selected-subs}

The analysis seeks to isolate a subsample from the CC inclusive sample which is enriched in CCQE events.   
This subsample, referred to as the CCQE enhanced sample, serves as the CCQE signal sample for determination
of the axial-vector mass and is presented in Sec.~\ref{sec:CCQE-sample}.    Before obtaining the signal
sample however, it is useful to elucidate the backgrounds that complicate the study of CCQE interactions. 
For these purposes, five mutually exclusive subsamples have been extracted from the CC inclusive sample.
Four of these serve as kinematic sideband samples to CCQE,
providing perspectives and constraints on background reaction 
categories, while the fifth subsample is the signal sample.   
An important development, described in
 Sec.~\ref{subsec:Suppression}, is the use of sideband samples 
 enriched in CC baryon resonance production 
to develop a data-driven correction to the $Q^2$ distribution 
of that background reaction category.

\subsection{Kinematic sideband samples}
\label{subsec:selected-sidebands}

Non-quasielastic reactions capable of mimicking the CCQE topology in the MINOS detector
consist of CC DIS events with low pion multiplicity and CC baryon resonance production events
($\mu^{-} + \Delta$ or $N^*$ final states).
The extent to which the {\small NEUGEN3} neutrino generator accurately describes these background
categories is investigated using four non-overlapping subsamples from the CC inclusive event sample.     
Their extraction from the inclusive CC sample is based
upon the hadronic mass $W$ as follows:

\begin{enumerate}

\item{\bf High-$Q^2$ DIS sample:}  Charged-current events of the 
deep inelastic scattering regime are isolated by requiring $W$ $>$ 2.0\,GeV. 
 Events having $W$ $>$ 2.0\,GeV and $Q^2$ $>$ 1.0\,GeV$^2$ 
 comprise the {\it  high-$Q^2$ DIS sample}.   
 According to the reference MC, this sample is completely dominated by true CC DIS events, 
 however there is also a few percent contribution from CC baryon resonance production.

\item{\bf Low-$Q^2$ DIS sample:}  Events having hadronic mass 
$W$ $>$ 2.0\,GeV and $Q^2$ $<$ 1.0\,GeV$^2$ comprise the {\it  low-$Q^2$ DIS sample}.
According to the MC, this sample is also dominated by CC DIS events, 
however the fraction of CC baryon resonance production events 
is larger ($\sim 10\%$) than is the case for the high-$Q^2$ DIS sample.

\item{\bf RES-to-DIS transition sample:}   Selection of events 
having hadronic invariant mass within the interval $1.3 < W < 2.0$\,GeV 
isolates a baryon resonance-to-DIS transition sample,
referred to hereafter as the {\it  RES-to-DIS transition sample}. 

\item{\bf QE-RES enriched and RES-enhanced samples:}  Selection of CC events having 
$W < 1.3$\,GeV yields a sample which 
is dominated by the quasielastic and baryon resonance production channels 
(hereafter, the {\it QE-RES enriched sample}). 
A cut on the energy of the hadronic shower recoiling from the muon, $E_{had}$,
is used to separate this sample into two subsamples, 
according to whether $E_{had}$ falls above or below 250 MeV.   
The subsample for which $E_{had}$ is greater than 250 MeV 
is referred to as the CC {\it RES-enhanced sample}. 
In the $Q^2$ region $0.0 < Q^{2} <$ 0.5\,GeV$^{2}$, the
reference MC predicts the latter sample
to be dominated by baryon resonance production 
with the $\Delta(1232)^{++}$ being 
the most abundant baryon resonance state.  This sample also contains a 
sizable CCQE component at moderate and high $Q^{2}$.
\end{enumerate}

\subsection{Sideband $Q^2$ distributions}
\label{subsec:sideband-Q2}

\subsubsection{Restriction to shape-only comparisons}
\label{subsubsec:Restrict}

The overall normalization of the absolute neutrino flux 
for this exposure is known to have an uncertainty of order 10\%. 
This systematic is dominated by uncertainties in the modeling 
of hadron production from the graphite target~\cite{Korzenev-Nu14, Harris-NuInt-14}.   
Additionally there are uncertainties associated with the shape of the flux spectrum 
in regions most relevant to this analysis.   To avoid these significant sources of error
and their complicated systematics, the analysis forgoes inferences
based upon differences in total event rate between data and the MC predictions.    
Rather, the approach taken is to restrict to shape-only comparisons, with emphasis 
placed upon the distributions of selected CC sideband and signal event
samples in four-momentum transfer, $Q^{2}$.    Consequently, in all subsequent
Figures showing MC comparisons to data, the MC prediction is shown scaled to 
the same number of events as in the data for the kinematic range displayed in each plot.
The scaling of the MC in this way is denoted in all cases by the plot interior label ``(Area Normalized)".
For subsamples in which the CCQE component is sizable (the QE-RES enriched, RES-enhanced, and CCQE enhanced samples),
 the scale factors (MC/data) which map POT-normalization into area-normalization fall within the range 1.08 to 1.19.  

\begin{figure}
\begin{center}
\includegraphics[width=8.5cm]{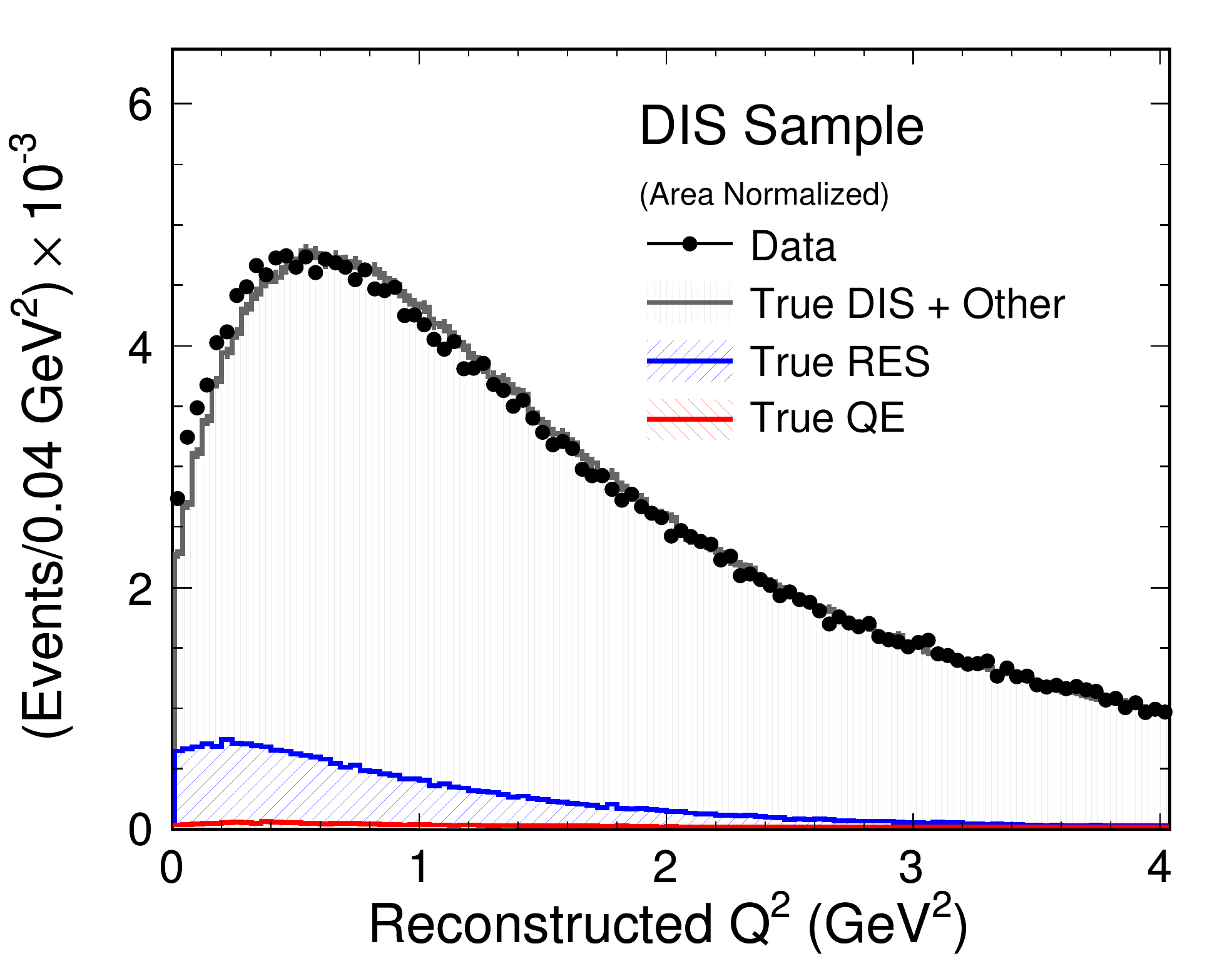}
{\caption{Combined distribution of reconstructed $Q^2$ for events of the 
high-$Q^2$ DIS and low-$Q^2$ DIS samples (distributions above, below 1.0\,GeV$^2$ respectively).   
The samples probe the MC model for DIS reactions of hadronic mass  $W > 2.0$\,GeV.  
The MC prediction (histograms, stacked) describes the general trend of the data (solid circles) but with
discrepancies which reflect uncertainties in parameters of the DIS model.
\label{fig.dissamples}
}}
\end{center}
\end{figure}
\begin{figure}
\begin{center}
\includegraphics[width=8.5cm]{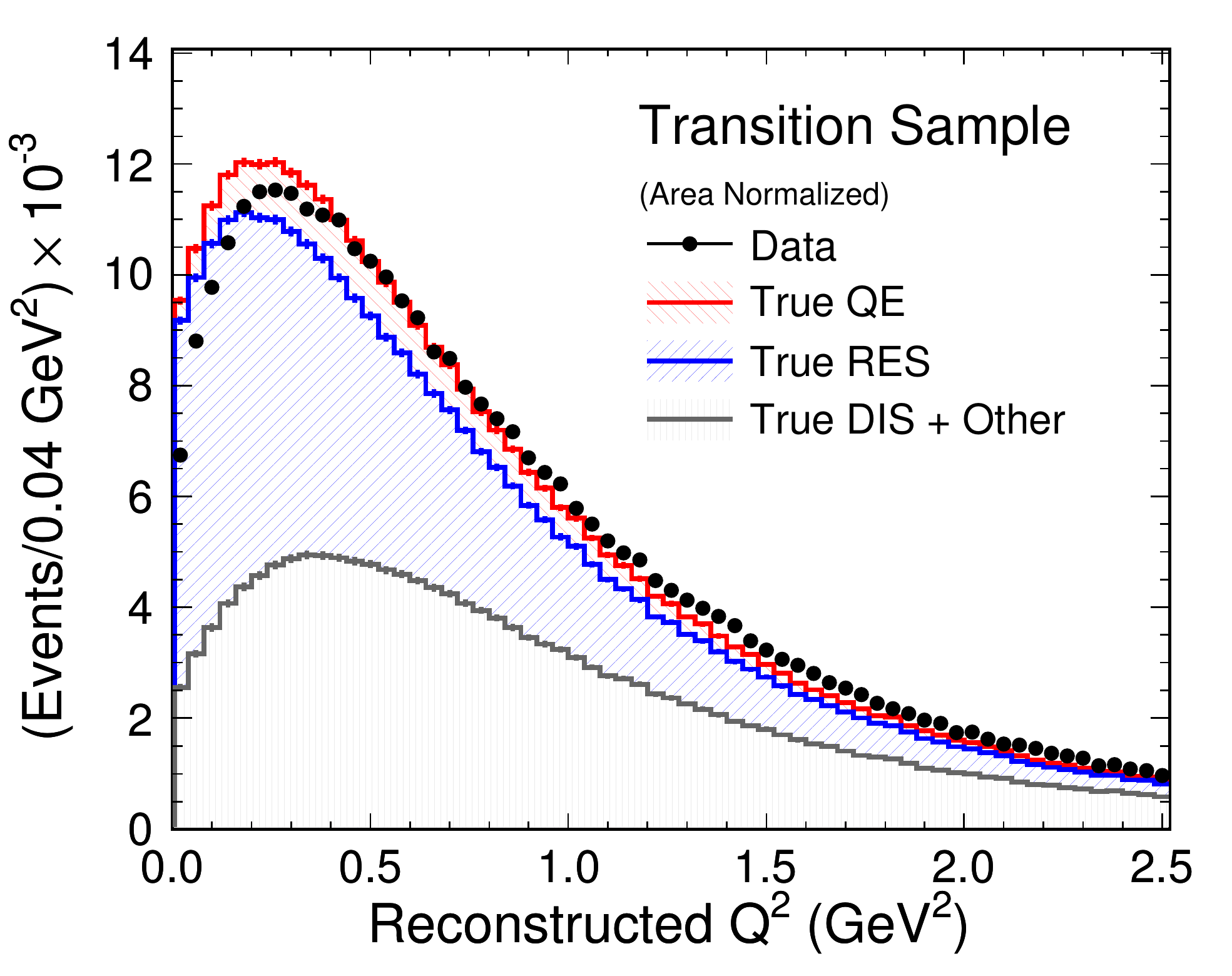}
{\caption{Distribution of reconstructed $Q^2$ for events of the 
 RES-to-DIS transition sample ($1.3 < W < 2.0$\,GeV).   The MC prediction (stacked histograms)
 is normalized to the number of data events.  
 The MC spectrum lies above the data
 over the low $Q^2$ region dominated by 
 CC baryon resonance channels.
 \label{fig.resdistransample}
}}
\end{center}
\end{figure}

\subsubsection{Data versus MC}
\label{subsubsec:side-compare}

The combined $Q^2$ distribution of the two DIS samples is displayed in Fig.~\ref{fig.dissamples}. 
The MC prediction is shown scaled to the same number of events as the data;
the MC/data scale factor in this case is 0.98.   This comparison checks the verity of the MC DIS model for
CC interactions with $W > 2.0$\,GeV, a region of hadronic mass lying well above the range $W < 1.3$\,GeV
from which CCQE candidate events are selected.    For the high-$Q^2$ DIS sample, the MC is observed to
match the data shape (and its absolute rate as well), for $Q^2$ from 2.0 to above 5.0\,GeV$^2$ (beyond the
range displayed in Fig.~\ref{fig.dissamples}).   Below 2.0\,GeV$^2$ and throughout the region of the low-$Q^2$
DIS sample, the MC describes the general trend of the data (histogram vs solid circles), however there are
discrepancies.   
These are indicative of shortcomings in the MC DIS model
which may affect the small DIS component ($\sim 11\%$) estimated to
reside in the selected signal sample.   They comprise a source of systematic uncertainty
whose presence is encompassed by error ranges 
allotted to parameters of the DIS model (Sec.~\ref{subsec:Systematic-errors}.6).

\begin{figure}
\begin{center}
\includegraphics[width=8.5cm]{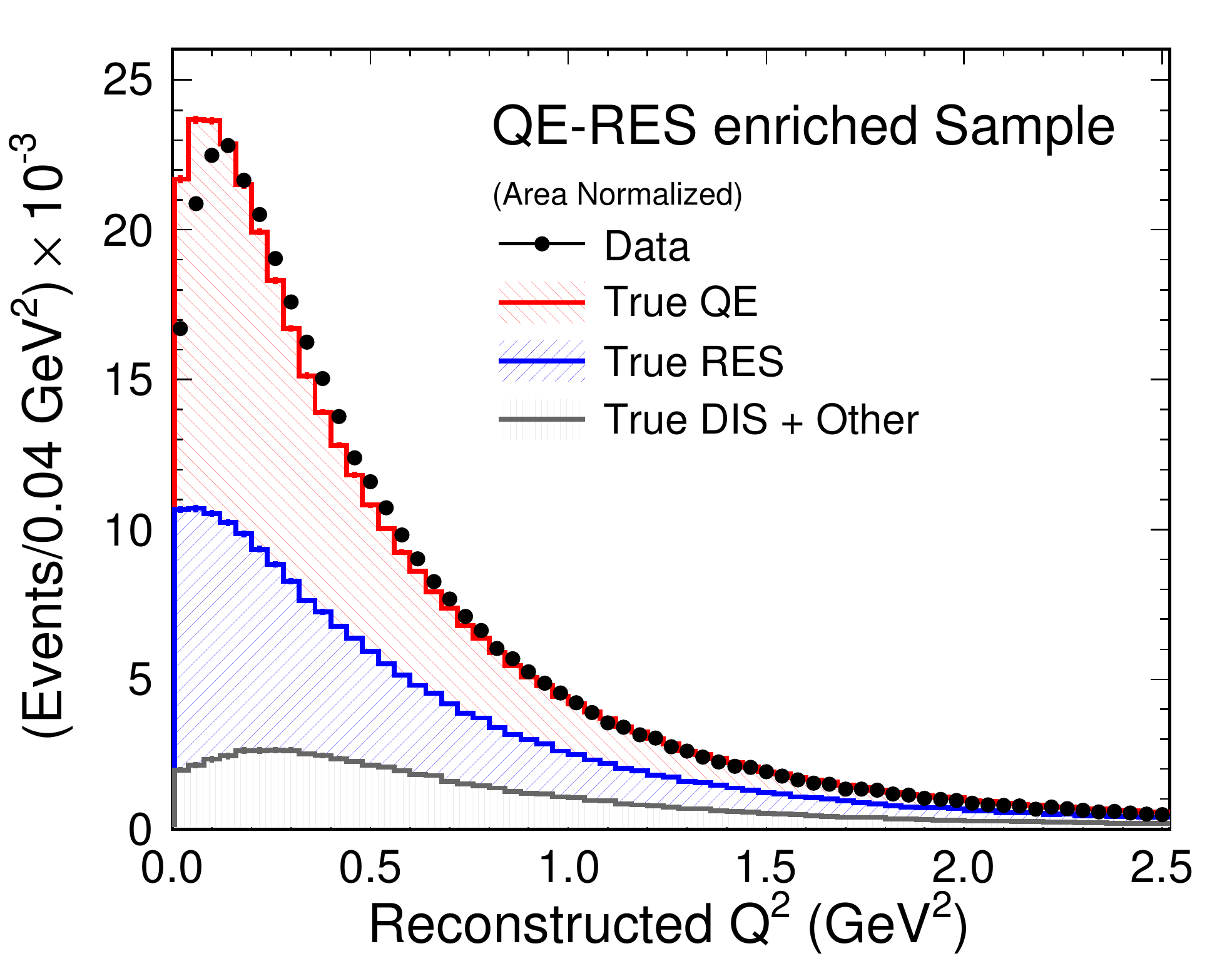}
\includegraphics[width=8.5cm]{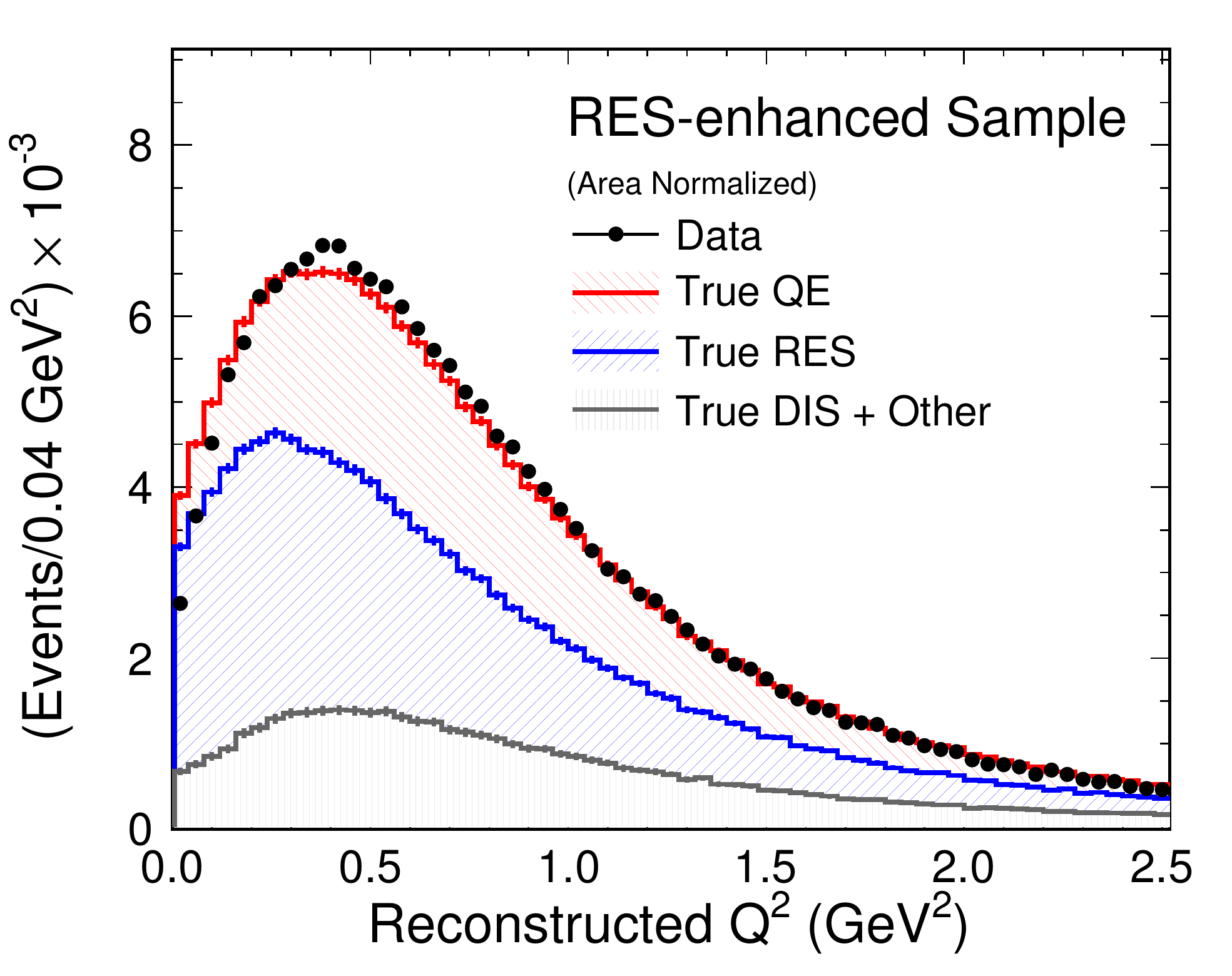}
{\caption{Distributions of reconstructed $Q^2$ for the 
 QE-RES enriched sample ($W < 1.3$\,GeV) and for
 the RES-enhanced sample  
 extracted from it by requiring $E_{had} > 250$\,MeV.   
 MC predictions are normalized to data event rates;
 the stacked histograms show contributions by reaction category.
 The distribution shapes in data versus MC show differences, particularly at $Q^2$ near 0\,GeV$^2$ 
 where CC baryon resonance channels (middle hatched regions) dominate the event rate.
 \label{fig.qeressamples}
}}
\end{center}
\end{figure}

The $Q^2$ distributions of data and MC for the 
RES-to-DIS transition sample are shown in Fig.~\ref{fig.resdistransample},
with the MC scaled to the total number of data events over the $Q^2$ range displayed.
The transition sample is predicted by the MC to be dominated 
by CC baryon resonance production throughout the low $Q^2$ region from 0 to $\sim$0.5\,GeV$^2$.
The MC exhibits an excess of event rate relative to the data throughout this low $Q^2$ region.

Selection of CC events having $W$ $<$ 1.3\,GeV yields 
the QE-RES enriched sample, from which the 
RES-enhanced sample is subsequently drawn.
The $Q^2$ distributions for the parent QE-RES enriched sample 
and for the `daughter' RES-enhanced sample are shown in the
upper and lower plots of Fig.~\ref{fig.qeressamples}, with the MC predictions (stacked histograms)
scaled to the number of data events in each plot.
From the component MC histograms it can be seen that
the DIS contribution is now smaller than the contributions from
CCQE events and from CC baryon resonance production.   
The RES-enhanced sample (lower plot) and the parent QE-RES enriched sample
as well (upper plot) possess a drop-off in rate at very low $Q^2$ that is not reproduced by the MC.

\subsection{Suppression of baryon resonances at low $Q^2$}
\label{subsec:Suppression}

The agreement of the MC with the $Q^2$ data distributions 
of the sideband samples, and of the signal-enhanced sample as well, is 
significantly improved by introduction of suppression of baryon resonance production at low $Q^2$.
(The DIS scattering model is not affected.)
A low-$Q^2$ suppression effect for CC two-body $\Delta(1232)$ production which extends 
beyond $Q^2 \sim 0.3$\,GeV$^2$ has been invoked in analyses of
the MiniBooNE data~\cite{MiniBooNEMA:2010, 
MiniBooNEPiplus:2011, MiniBooNEPizero:2011}. (See also Ref.~\cite{ref:Paschos-NPRS-2005}).
The proposed effect resembles the low-$Q^2$ suppression exhibited by treatments that go
beyond the Fermi gas model, such as the Random Phase Approximation (RPA)~\cite{Marteau:1999kt, ref:GandS-2003, Nieves:2004wx, Martini:2009uj}, 
nuclear spectral functions~\cite{BenharMeloni:2009}, or the relativistic distorted-wave impulse approximation (RDWIA) 
as calculated for CCQE interactions~\cite{Butkevich:2008}.
Since the $\Delta(1232)$ and higher mass baryon resonance states are often too short-lived to escape the parent
nucleus before decaying, Pauli blocking may account for part of this effect.
As discussed below, the analysis finds that a suitable suppression factor is one that removes about
20\% of two-body CC $\Delta$/N$^{*}$ production in the MC model.   
Of course the introduction of a suppression factor to be included 
in the MC model prediction has its own sources of uncertainty;  
these are accounted for in the error treatment of this analysis.

The RES-enhanced and RES-to-DIS transition samples were fitted together over the range 
0 $\leq$ $Q^2$ $<$ 0.6\,GeV$^2$ using a resonance suppression described as a function of true $Q^{2}$. 
The motivation is that, within this $Q^{2}$ range, baryon resonance production 
is the dominant reaction category in each sample.   As described below, a similar functional shape is found
to describe the suppression at low $Q^{2}$ in both samples.

 The fitting to the two sideband samples was carried out as a multi-step process.  At the outset a
 candidate shape for weighting to be applied in bins of true $Q^{2}$ was specified.   
 The predicted contribution to each bin was then adjusted, one bin at a time, 
 to a value that reduced the residuals over the two reconstructed samples 
 after the samples were area-normalized.   This yielded a suppression shape that better described 
 the data.   The next step was to fit an overall strength parameter in conjunction with the refined shape. 
 The procedure for the two previous steps was then iterated and the change to the suppression
 parametrization was found to be negligible.

A systematic error band was constructed by evaluating the effects of error sources
expected to be significant for the $M_A^{QE}$ measurement.
For $Q^2$ values below ~0.3\,GeV$^{2}$ the shape of the error band reflects 
uncertainties arising from the muon and hadron energy scales and from the intranuclear rescattering model,
which affect event selection and Q$^2$ reconstruction.  Also included is an allowance for sensitivity to 
higher-than-nominal $M_A^{QE}$ values of the magnitude determined in this work.   This sensitivity enters
the construction of the suppression weight through the presence of CCQE background events in the sideband samples.
At higher Q$^2$ the error band reflects
uncertainty in the effective turnoff point for the suppression~\cite{ResSuppNote}.

\begin{figure}
\begin{center}
\includegraphics[width=8.0cm]{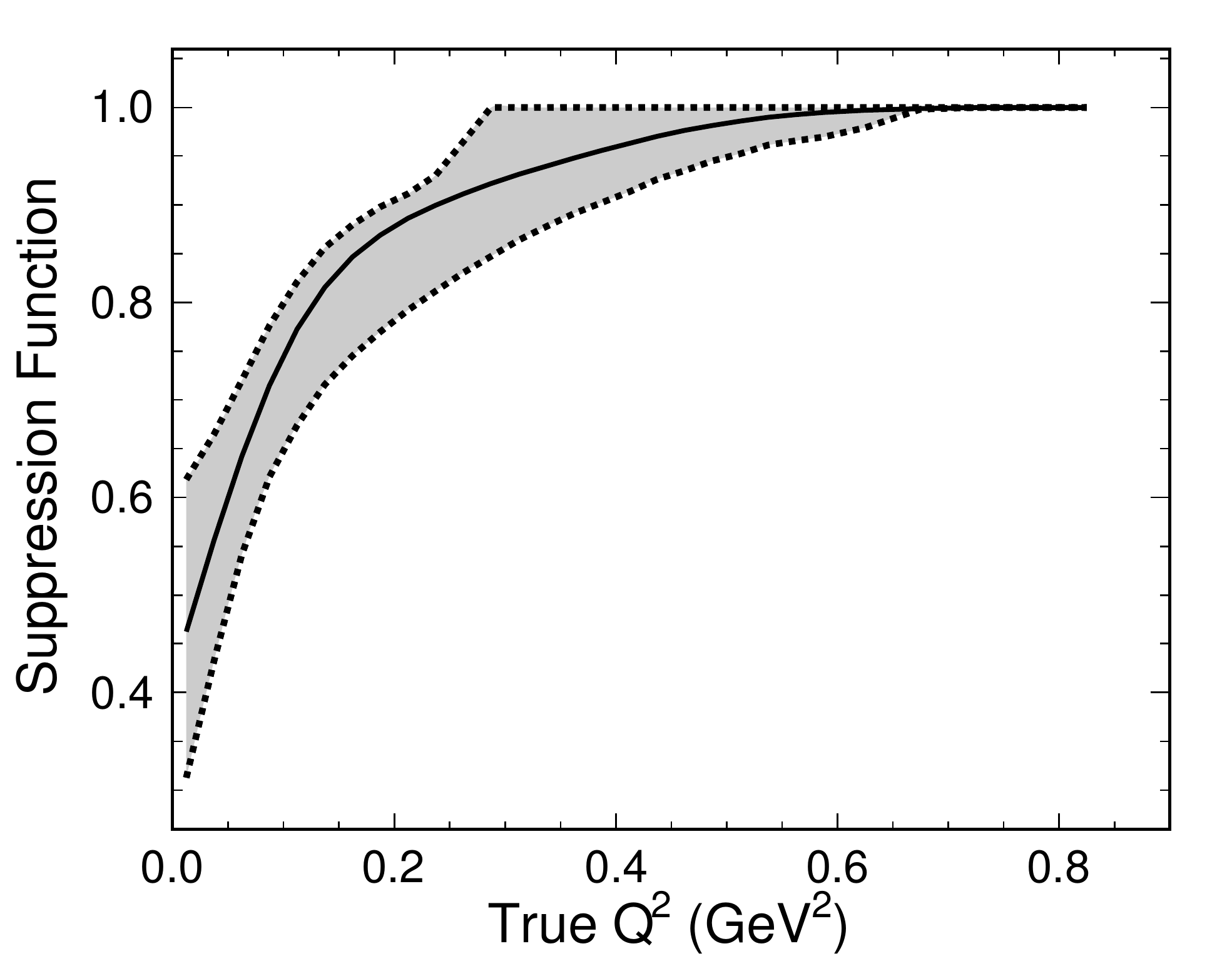}
{\caption{The $Q^2$-dependent weight function which, when
applied to the MC model of baryon resonance production, brings the MINOS MC predictions into agreement
with data for sideband samples dominated by $\Delta$/N$^{*}$  production.    The shape
and strength of the suppression are sensitive to systematic uncertainties as indicated by the error band.
\label{fig.errorband}
}}
\end{center}
\end{figure}

The suppression function with its error band is shown in Fig.~\ref{fig.errorband}.   
The function is applied to true baryon resonance events 
generated by the MC that enter the background estimate for the $M_{A}^{QE}$ measurement. 
(For $0.0 < Q^2 < 0.7$\,GeV$^2$, the central curve of Fig.~\ref{fig.errorband} is replicated by the phenomenological form:
$f(Q^2) = A \times [ 1 + {\rm exp}\{ 1 - \sqrt{Q^2}/Q_{0}\}]^{-1}$, with $A = 1.010$ and $Q_{0} = 0.156$\,GeV.)
This data-driven suppression function remedies the discrepancy 
in the very low-$Q^2$ spectra of both the RES-enhanced and the RES-to-DIS transition sideband samples, 
two completely independent samples which contain very different admixtures of background processes.

\section{CCQE enhanced sample}
\label{sec:CCQE-sample}

\subsection{Event selection; CCQE kinematics}
\label{subsec:Select-CCQE}

The final subsample to be drawn from the CC inclusive sample, 
one which has no overlap with the sideband samples, is 
the CCQE enhanced sample.   Such a sample, enriched in signal events, 
is isolated by exploiting the tendency
of CCQE interactions to deposit relatively small amounts of hadronic energy 
in the MINOS detector as illustrated by Fig.~\ref{fig:ccSample_eshw}.
The topology targeted is a single muon track, either with 
no additional energy deposition in the event or else with an 
accompanying hadronic system having $E_{had}$ less than a few hundred MeV.
Three criteria are used to select the candidate signal events of the CCQE enhanced sample:

\begin{enumerate}
\item  A selected event contains a single muon track 
           (in accord with the criteria of Sec.~\ref{subsec:select-cc-inclusive}).
\item  The single reconstructed track is required to stop in the detector 
            and not on the far side of the magnetic coil.  The muon
           end point in alternate view planes 
           must be separated by $\le$ 5 planes (15 planes) for the calorimeter (spectrometer).
           At the far end of the spectrometer, the endpoint must be contained by at least two tracking planes.        
\item  The reconstructed final-state hadronic system is required to have energy, $E_{had}$, less than a designated 
           threshold value.    As indicated by the distributions 
           in Fig.~\ref{fig:ccSample_eshw}, the threshold values of interest lie
           in the range $0 < E_{had} <  500$ MeV.   
           Based upon considerations of CCQE sample purity and the efficiency for retaining
           signal events, the selection threshold requirement for $E_{had}$ is set to  $E_{had} <$ 225 MeV.    
\end{enumerate}

\noindent  The second criterion is motivated by the fact that in the MINOS Near Detector, 
determination of muon momentum by range yields 
a more accurate and higher resolution measurement
than does measurement based upon track curvature.    
The second and third criteria constrain the kinematic distributions of the selected CCQE sample.
The requirement that final-state muons stop in the detector effectively limits the sample to events with $E_{\nu} < 8$\,GeV;
the $E_{had}$ restriction improves the sample purity but also removes genuine quasielastic events with large $Q^{2}$. 

Reconstruction of muon momentum with good angular and momentum resolution 
is important because analysis of the CCQE enhanced sample can 
utilize a reconstruction of $Q^2$ based upon the QE hypothesis 
and muon kinematics, rather than relying on hadronic
calorimetry as is done for the CC inclusive and the kinematic sideband samples.
The neutrino energy and $Q^{2}$ can be estimated event-by-event
by using the reconstructed muon track, under the assumption
that each event is in fact a CCQE scatter from a stationary bound neutron.    
The expressions for these quantities, designated as $E_{\nu}^{QE}$ and $Q^{2}_{QE}$ are:
\begin{equation}\label{eq:E-QE}
E_{\nu}^{QE} \equiv \frac{(M_{n}-\epsilon_{B})E_{\mu}+(2M_{n}\epsilon_{B}-\epsilon_{B}^{2}-m_{\mu}^{2})/2}
{(M_{n}-\epsilon_{B})-E_{\mu}+p_{\mu}\cos\theta_{\mu}}~,
\end{equation}
\noindent and
\begin{equation}\label{eq:Q2-QE}
Q^{2}_{QE} \equiv 2E_{\nu}^{QE}(E_{\mu}-p_{\mu}\cos\theta_{\mu})-m_{\mu}^{2}~.
\end{equation}
\noindent   For the reconstructed neutrino energy, $E_{\nu}^{QE}$, the parameter $\epsilon_{B} = +34$ MeV 
accounts for the nucleon binding energy, or the average nucleon removal energy, of the target neutron within the iron nucleus.
As an estimator of $Q^2$, $Q^{2}_{QE}$ of Eq.~\eqref{eq:Q2-QE} is unbiased for genuine CCQE events, but is biased towards
lower values for the baryon resonance background reactions.

\begin{figure}
\begin{center}
\includegraphics[width=8.5cm]{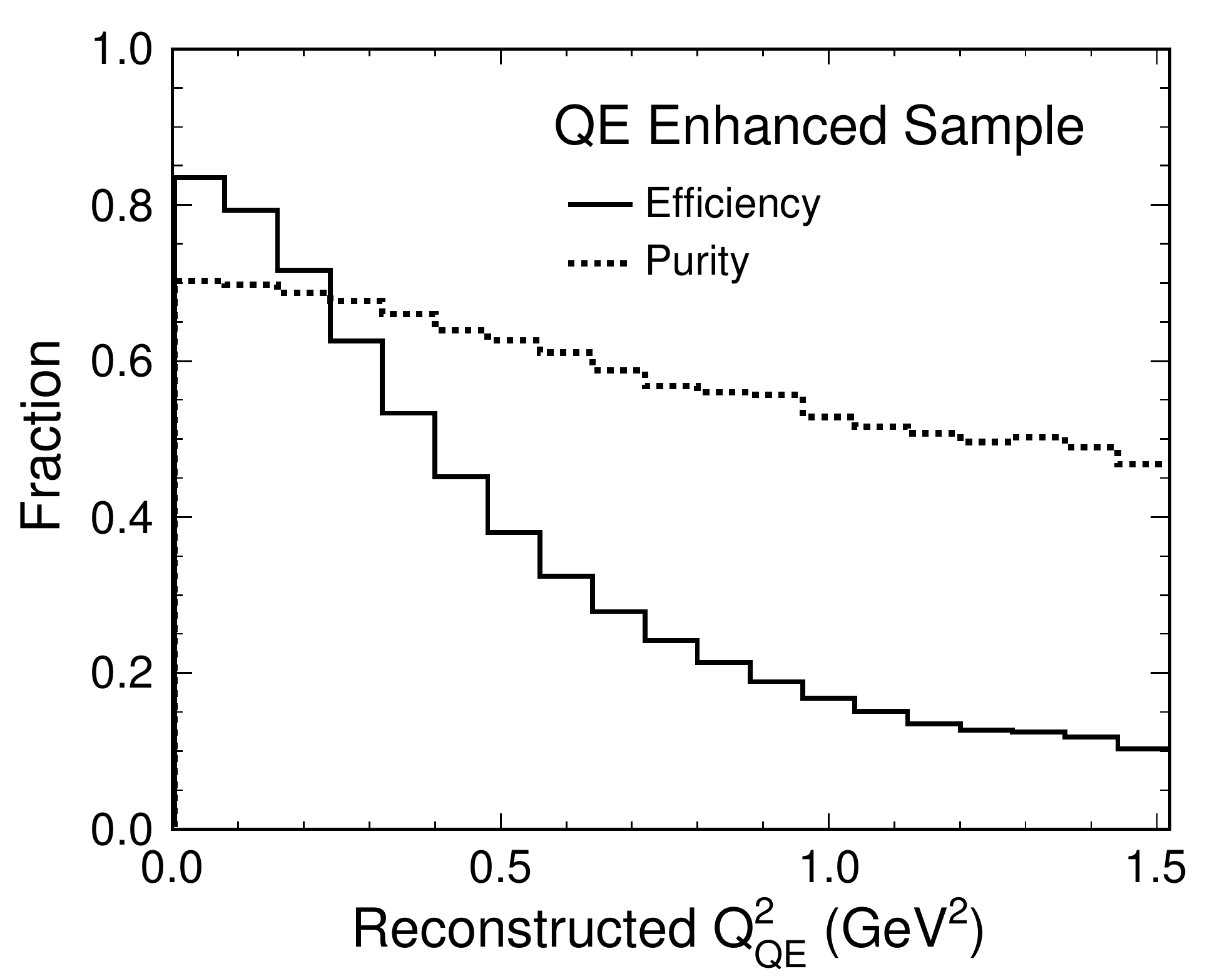}
\caption{The selection efficiency (solid-line histogram) and the purity (dotted-line histogram)
for CCQE candidate events extracted from the CC inclusive sample, as a function of reconstructed $Q^{2}$.
The fall-off of efficiency with increasing $Q^{2}_{QE}$ is a consequence of the restriction 
to low $E_{had}$ events.
 \label{fig:qeSample_effPurVsQ2}}
\end{center}
\end{figure}

\subsection{Selection efficiency and sample purity}
\label{subsec:efficiency-purity}

Figure \ref{fig:qeSample_effPurVsQ2} 
shows the efficiency (solid-line histogram) and 
sample purity (dotted-line histogram) for the selected CCQE enhanced sample 
as a function of reconstructed $Q^{2}_{QE}$.    The efficiency is highest
at $Q^{2}_{QE}\simeq 0.0$; the residual $16\%$ of inefficiency at $Q^{2}_{QE}\simeq 0.0$ arises primarily from
the muon containment requirement.   The gradual fall-off of efficiency with increasing $Q^{2}_{QE}$ is 
due to the restriction imposed on the energy $E_{had}$ of the recoiling hadronic system.
In recognition of efficiency reduction at high $Q^{2}_{QE}$,  
the fitting of the CCQE enhanced sample for  $M_{A}$ is 
restricted to events having reconstructed $Q^2_{QE}$ in the range $0.0 < Q^{2}_{QE} < 1.2$\,GeV$^{2}$.
The purity of the CCQE selected subsample exceeds 50\% 
for all reconstructed $Q^{2}_{QE}$ values below 1.2\,GeV$^{2}$.

Event statistics for the CCQE enhanced sample together with an estimate of its reaction
composition are presented in Table~\ref{tab.SelectedEvents}.   
The populations of component CC reaction categories according to the MC model are
tabulated in the upper rows.   The lower rows show the MC predictions for the 
data exposure together with the numbers of data events.  
Also shown in the rightmost column are corresponding breakouts by reaction type, rates and ratios
for the sample restricted by a selection ($E_{\nu} < 6.0$\,GeV) which removes the high-$E_{\nu}$ tail
of events.   The data-over-MC ratio (bottom row) for either the full or restricted signal sample
shows the observed candidate event rate to exceed the MC prediction by $19\%$.

\begin{table}
\begin{center}
\begin{tabular}{lrrr}
\hline
\hline
\multicolumn{3}{c}{\rule{0pt}{2.4ex} CCQE Enhanced Sample Composition} \\
\hline
\rule{0pt}{2.4ex} MC Reaction Type & ~All $E_{\nu}$ & $~~E_{\nu} \leq$ 6 GeV \\
\hline
$\nu_{\mu}$-CC QE  &  123,310 & 120,820\\
$\nu_{\mu}$-CC RES &  41,060  & 40,110 \\
$\nu_{\mu}$-CC DIS &  21,260  & 20,580 \\
$\nu_{\mu}$-CC COH &   370   & 360  \\
$\nu_{\mu}$-NC     & 420   & 420  \\
$\bar{\nu}_{\mu}$  &   110    & 110    \\
\hline
\rule{0pt}{2.4ex} Total MC                      & 186,530 & 182,400\\
\hline
\hline
\rule{0pt}{2.4ex} Data                    & 221,300 & 216,560\\
\hline
\rule{0pt}{2.4ex} Data/MC Ratio & 1.186 & 1.187 \\ 
\hline
\hline
\end{tabular}
{\caption{Event populations for the CCQE enhanced sample, in the MC model and in the data.
Upper rows show the sample composition by reaction category as estimated by
the MC model for the $1.26 \times 10^{20}$ POT data exposure.   Comparisons of the numbers of CCQE candidate events as 
predicted by the MC versus the numbers of data events are provided by the lower rows.
\label{tab.SelectedEvents}}}
\end{center}
\end{table}

In Table~\ref{tab.SelectedEvents} and throughout this work,  MC processes are labeled according to
the interaction type that is `as born'  inside the target nucleus.   Thus signal events are events that originate
as QE according to the MC, and the number of such events in the data is inferred from the MC.
The topologies that emerge from the struck nucleus however, are subject to alterations by final state
interactions.   Among the as-born baryon resonance events (as-born DIS events), 28\% (21\%) are devoid of pions upon
exiting the struck nucleus.    These backgrounds are among the 73\% of events in 
the simulated CCQE enhanced sample for which the final state released from the target nucleus consists solely of
a muon plus nucleon(s).

\subsection{Sample $E_{\nu}^{QE}$ and $Q^{2}_{QE}$ distributions}
\label{subsec:CCQE-sample}

\begin{figure}
\begin{center}
\includegraphics[width=8.5cm]{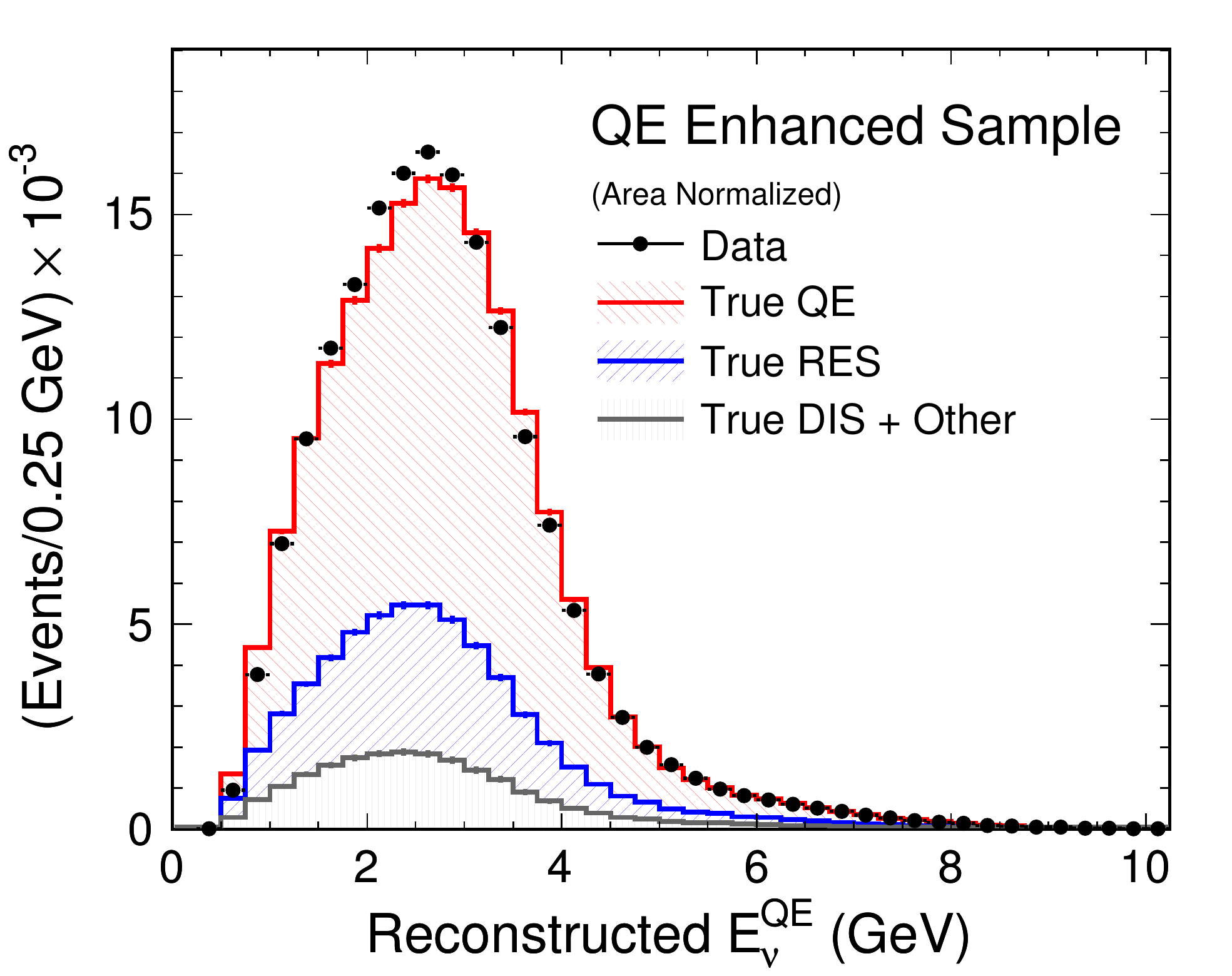}
\includegraphics[width=8.5cm]{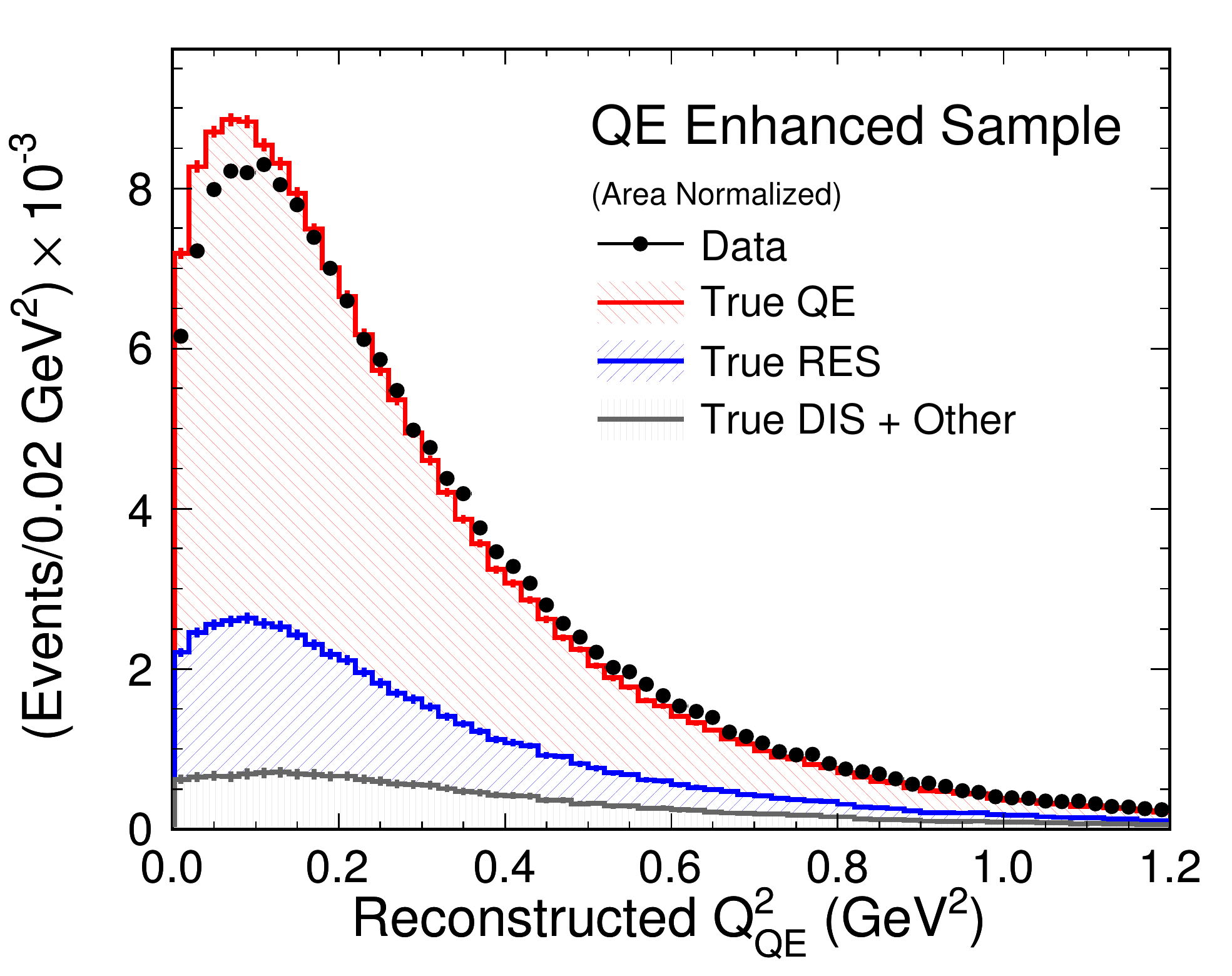}
{\caption{Distributions of reconstructed neutrino energy, $E_{\nu}^{QE}$ (top), 
and of $Q^{2}_{QE}$ (bottom) for CCQE selected data (solid circles) and 
for the flux-tuned MC (stacked histograms).   The MC prediction includes
the data-driven suppression weighting for CC baryon resonance production shown in
Fig.~\ref{fig.errorband}.   The MC distributions are scaled to match the data rate for
$Q^{2}_{QE}$ below 1.2\,GeV$^{2}$. 
In the lower plot, the MC is observed to exceed (fall below) the data for $Q^{2}_{QE}$ less than
(greater than) 0.2\,GeV$^2$.   
 \label{fig.qeSampleEnuAndQ2}}}
\end{center}
\end{figure}

Comparisons of MC predictions to data are shown in Fig.~\ref{fig.qeSampleEnuAndQ2}
for $E_{\nu}^{QE}$ and $Q^{2}_{QE}$ distributions of
the CCQE-enhanced sample; as with the sideband samples,
the MC is plotted area-normalized to the data.  
The hatched component histograms show the extent 
to which the CCQE signal is expected to dominate the sample.
In this figure and in subsequent comparisons,
the suppression of baryon resonance production 
at low $Q^2$ is incorporated into the MC prediction 
as described in Sec.~\ref{subsec:Suppression}.    

The analysis now focuses upon the $Q^{2}_{QE}$ distribution
of Fig.~\ref{fig.qeSampleEnuAndQ2} (bottom); the remaining data-vs-MC discrepancies are
to be accounted for by fitting model parameters that alter the MC $Q^{2}$ distribution so as to better describe the data.
The MC prediction (histogram, area-normalized) is observed to exceed the data
(solid circles) in the region $ 0.0 < Q^{2}_{QE} < 0.2$\,GeV$^2$, and to fall below the data in all bins of the
higher range $Q^{2}_{QE} > 0.2$\,GeV$^2$.   The fit is capable of addressing these differences by determining 
the value of the axial mass $M_{A}^{QE}$ that yields the best match 
of the MC to the data over the $Q^{2}_{QE}$ range shown in Fig.~\ref{fig.qeSampleEnuAndQ2}.
The discrepancy in the very low-$Q^2$ region indicates that
the amount of Pauli blocking for CCQE events 
(governed by the $k_{Fermi}$ parameter) is to be increased,
while the differences at higher $Q^{2}_{QE}$ 
suggest that $M_{A}^{QE}$ values above 1.0\,GeV are to be favored.

\section{Determination of Effective $M_{A}$ }
\label{sec:Effective-MA}

As previously noted, the analysis foregoes the use of absolute event rate information.   Rather,
measurement of the effective $M_{A}$ for quasielastic scattering in iron 
is based on the shape of the distribution of candidate CCQE events in
the variable $Q^{2}_{QE}$.

\subsection{Fit procedure}
\label{subsec:Fit-Procedure}

With the suppression weight now included in the MC modeling 
of CC baryon resonance production, 
the analysis focuses on the CCQE-enhanced sample and its
distribution in reconstructed $Q^{2}_{QE}$, shown 
in the lower plot of Fig.~\ref{fig.qeSampleEnuAndQ2}.
The modified MC prediction, with the axial-vector mass $M_{A}$ 
treated as a free parameter, is to be fitted to the data.
The fit is carried out by minimizing the following $\chi^2$:
\begin{equation}\label{eq.chi2}
\chi^{2} = \sum_{i=1}^{N_{bins}}\frac{(N_{i}^{obs}-N_{i}^{MC}(M_{A}, \alpha_{j = 1,3}))^{2}}{(N_{i}^{obs}+
r_{0} \cdot N_{i}^{MC}(M_{A}, \alpha_{j = 1,3}))}
+ \sum_{k=1}^{3}\frac{(\Delta\alpha_{k})^{2}}{\sigma_{\alpha_{k}}^{2}}.
\end{equation}
\noindent Here, $N_{i}^{obs}$ is the observed number of events in data for bin $i$, 
and $N_{i}^{MC}(M_{A}, \alpha_{1}, \alpha_{2},\alpha_{3})$ is the number of events 
predicted by the MC using the current values of the fit parameter $M_A$ and
the three nuisance parameters, $\alpha_{j}$ for j = 1,2,3. 
The constant $r_{0}$ in the denominator is the ratio of POT in the data to POT in the MC.
The MC prediction $N_{i}^{MC}$ also contains a scale factor which sets the number of MC interactions
equal to the number of data interactions.   The latter factor is computed at the beginning of every trial fit
and reduces the fit degrees of freedom by one.  Thus as the parameters change the number of MC events,
the $\chi^2$ evaluates the match to the shape of the data $Q^2$ distribution.
The rightmost summation is over the penalty terms, each of which 
is the square of $\Delta\alpha_{k}$, the shift from nominal for 
the $k^{th}$ systematic parameter, divided by the square of $\sigma_{\alpha_{k}}$, the $1\sigma$ error assigned to 
the $k^{th}$ systematic parameter.

The principal fit uses four parameters.    
The axial-vector mass, $M_{A}$, is the single free parameter.   
It is fitted in conjunction with three nuisance parameters:  
({\it i\,}) A scale parameter for the momentum assignment to stopping muons
for which $\pm 1\sigma$ corresponds to $\pm 2$\% ~\cite{prd_1e20};  
({\it ii\,})  the axial-vector mass for CC baryon resonance production, $M_{A}^{RES}$,
having nominal value 1.12\,GeV with uncertainty (at 1 $\sigma$) of $\pm 15\%$~\cite{prd_1e20, Kuzmin:2006dh};
and ({\it iii\,})  the Fermi momentum
cutoff, $k_{Fermi}$, whose value (263 MeV/c for neutrons in iron) is used by {\small NEUGEN3} to set the strength 
of Pauli blocking for CCQE interactions within target nuclei.  

\begin{figure}
\begin{center}
\includegraphics[width=8.5cm]{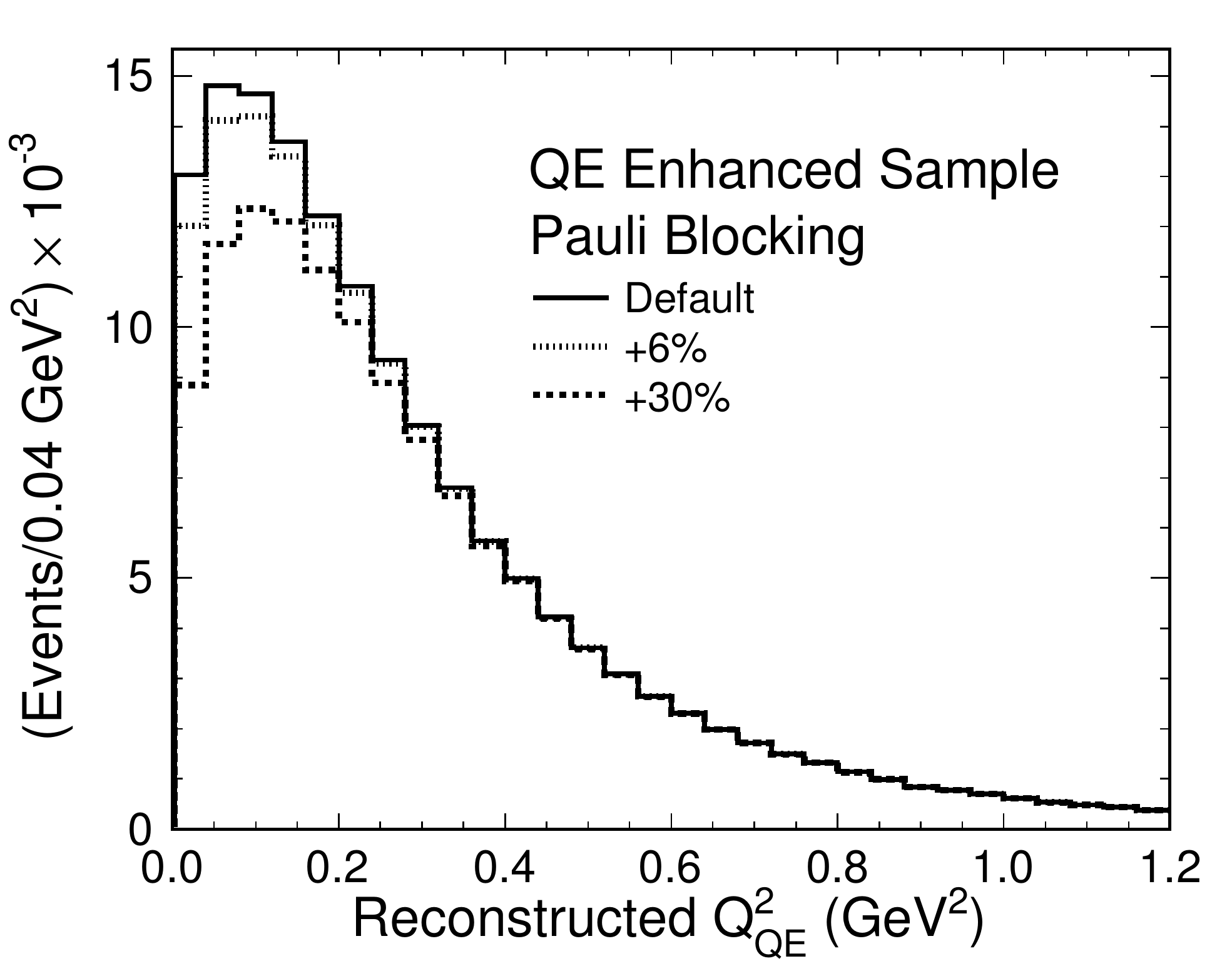}
\caption{Enhanced suppression of the MC $Q^{2}_{QE}$ distribution for the CCQE enhanced sample resulting from
increase in the upper momentum cutoff for nucleons in iron, $k_{Fermi}$, above its nominal value of 263 MeV/c.  
The  bold-dash histogram shows the effect of setting $k_{Fermi}$ at the upper bound of plausible values.
The principal fit of this analysis favors the milder suppression shown by the fine-dashed histogram.
\label{fig:lowQ2Suppression}}
\end{center}
\end{figure}

The $k_{Fermi}$ cutoff acts as an effective low-$Q^2_{QE}$ suppression parameter.  
It serves the same purpose as the $\kappa$ parameter 
used by MiniBooNE~\cite{MiniBooNEMA:2008,MiniBooNEMA:2010}.   
The parameter provides the fit with a proxy equivalent for a treatment of
Pauli blocking plus other nuclear effects that are operative at low $Q^2_{QE}$.
In {\small NEUGEN3}, the CCQE kinematics are computed for all possible four-momentum 
transfers.   However generated events having recoil nucleon momenta below the $k_{Fermi}$ limit are rejected.
Based on comparisons with models using nuclear spectral functions~\cite{BenharMeloni:2009} or using
RDWIA~\cite{Butkevich:2008}, additional amounts of suppression produced by increasing $k_{Fermi}$
by as much as 30\% are possible in theory.    The viability of this elevated parameter 
range is also supported by comparisons to current models with
RPA effects~\cite{Martini:2009uj, Nieves:2011yp, Martini:2011wp, Gran:2013kda}, and 
by the resonance suppression results described in Sec.~\ref{subsec:Suppression}.  
For these reasons  $k_{Fermi}$ is allowed to vary above its nominal in accordance with a 1 $\sigma$
uncertainty of 30\% during iterations of the principal fit.  
The range of low-$Q^2_{QE}$ suppression in CCQE accessible 
via $k_{Fermi}$ is illustrated by the lowest (bold-dash) histogram 
in Fig.~\ref{fig:lowQ2Suppression}.   As it turns out (see paragraphs below), the 
principal fit requires a relatively small amount of additional suppression, from a $k_{Fermi}$ increase of +6\%,
to describe the data.   The low-$Q^2$ suppression thereby implied to the $Q^{2}_{QE}$ distribution of the 
CCQE enhanced sample is shown in Fig.~\ref{fig:lowQ2Suppression} by the fine-dashed histogram.

\begin{table}
\begin{center}
\begin{tabular}{cccccccc}
\hline
\hline
\rule{0pt}{2.5ex} $Q^{2}_{QE}$ Range  & $M_{A}$ & $E_{\mu}$ & \small{$~M_{A}^{RES}$} & $k_{Fermi}$  \\
\rule{0pt}{2.2ex}~(GeV$^{2}$) & (GeV) & scale & (GeV) & scale  \\ 
\cline{1-5} 
&&&&\\ [-7pt]
0.0 - 1.2 & 1.23$^{+0.13} _{-0.09}$& ~1.00$\pm 0.01$ &~1.09$^{+0.14} _{-0.15}$ & 1.06$\pm 0.02$  \\ [4pt]
0.3 - 1.2 & 1.22$^{+0.18} _{-0.11}$ & ~1.00$^{+0.01} _{-0.02}$ &~1.09$^{+0.15} _{-0.16}$ & N.A. \\ [3pt]
\hline
\hline
\end{tabular}
\caption{Results from shape-only fits to the $Q^2_{QE}$ distribution
of the selected CCQE sample, for $M_{A}$, for the three nuisance
parameters, and the MC to data normalization obtained with the best fit parameters.
The fit over the full reconstructed $Q^2_{QE}$ range (upper row)
is compared to a fit in which the $Q^2_{QE}$ region 
most susceptible to nuclear distortions is left out (lower row).  }
\label{tab:fitResultsTab}
\end{center}
\end{table}

\subsection{Fitting the shape of the $Q^{2}_{QE}$ distribution}
\label{subsec:Fit-to-shape}

Since the MC versus data comparisons of Sec.~\ref{subsec:CCQE-sample} 
have shown the low-$Q^2_{QE}$ regime to be poorly modeled, the 
fitting of the reconstructed $Q^{2}_{QE}$ distribution of the CCQE-enriched sample was carried out 
using two different configurations.   For each configuration, the fitting of the augmented MC prediction to the data
is only for the shape of the $Q^{2}_{QE}$ distribution; 
an upper bound of 1.2\,GeV$^{2}$ is imposed on $Q^{2}_{QE}$ for events of either fit.    

In the principal fit, all events having reconstructed $Q^{2}_{QE}$ less than 1.2\,GeV$^{2}$ were 
included and the $k_{Fermi}$ parameter was allowed to vary.
The best-fit values thereby obtained are given in the upper row
of Table~\ref{tab:fitResultsTab}.   The principal fit yields a reduced $\chi^{2}$ per degree of freedom of 0.79; 
the uncertainty on the best-fit $M_{A}$ value due solely to statistical effects is $\pm$0.07\,GeV.

 For the alternate configuration,  only CCQE candidates having reconstructed $Q^{2}_{QE}$ 
values between 0.3\,GeV$^{2}$ and 1.2\,GeV$^{2}$ were used, and the normalization of MC events
to data was restricted to this reduced $Q^2$ range.   Furthermore, 
the $k_{Fermi}$ parameter was fixed at its nominal value.  
 As in all previous fit trials, low $Q^{2}$ suppression
of CC baryon resonance production is operative in the MC model.
A good fit to the data is obtained, indicating that the modeling augmentations at low $Q^{2}$ contribute to the 
agreement between MC and data obtained by this more restricted fit.    The values 
for the axial-vector mass, $M_{A}$, and for the three nuisance parameters describing the systematics
are shown in the lower row of Table~\ref{tab:fitResultsTab}; these are in excellent agreement 
with the results of the principal fit.

Concerning the absolute rate of events in the CCQE enhanced sample (not used in the fits),
the fit results imply $[N_{data}/N_{MC}]_{\small{CCQE}}$ = 1.09.    This value is an improvement
compared to the ratio 1.19 predicted by the original MC model (see Table~\ref{tab.SelectedEvents}).

\begin{figure}
\begin{center}
\includegraphics[width=8.8cm]{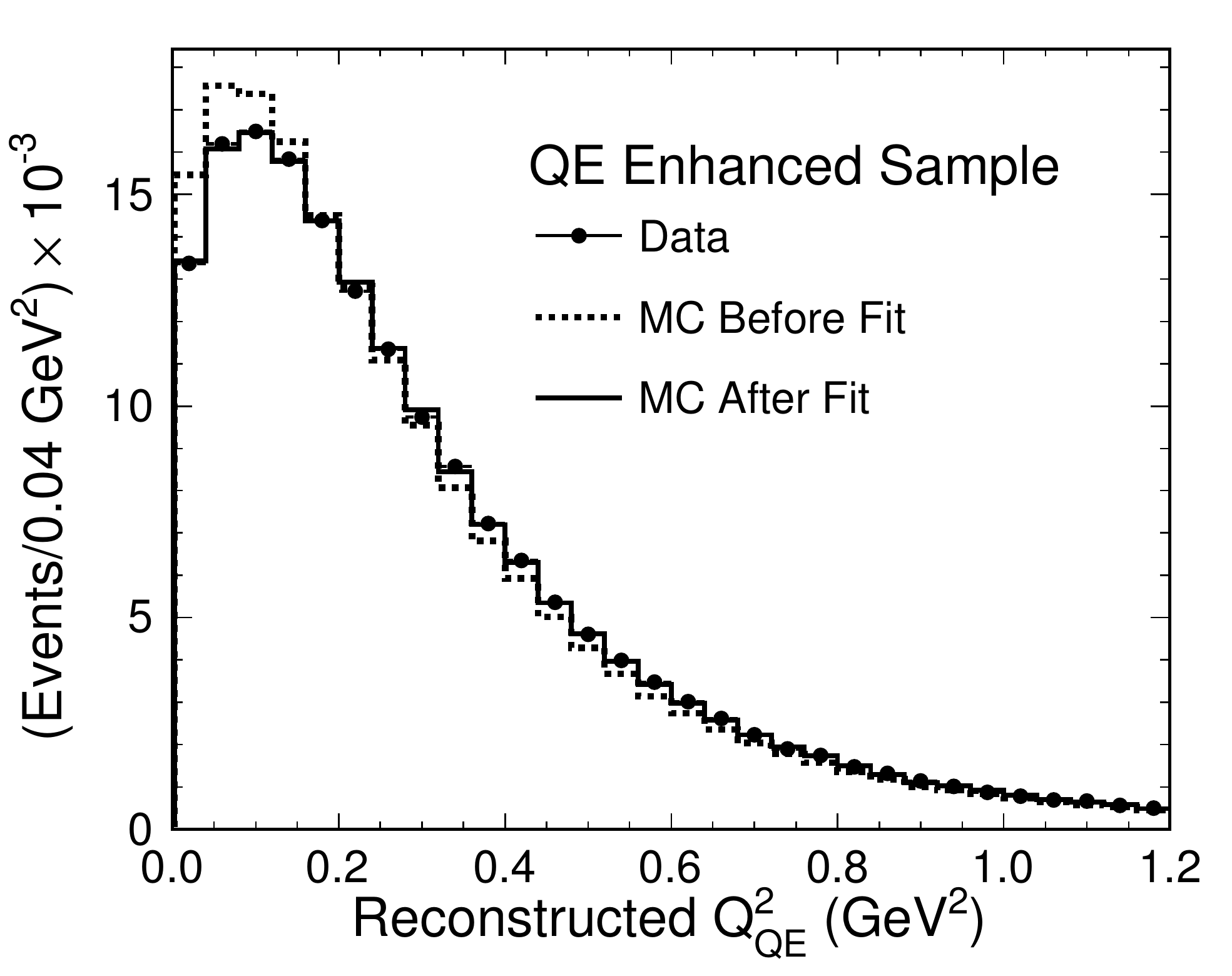}
\includegraphics[width=8.8cm]{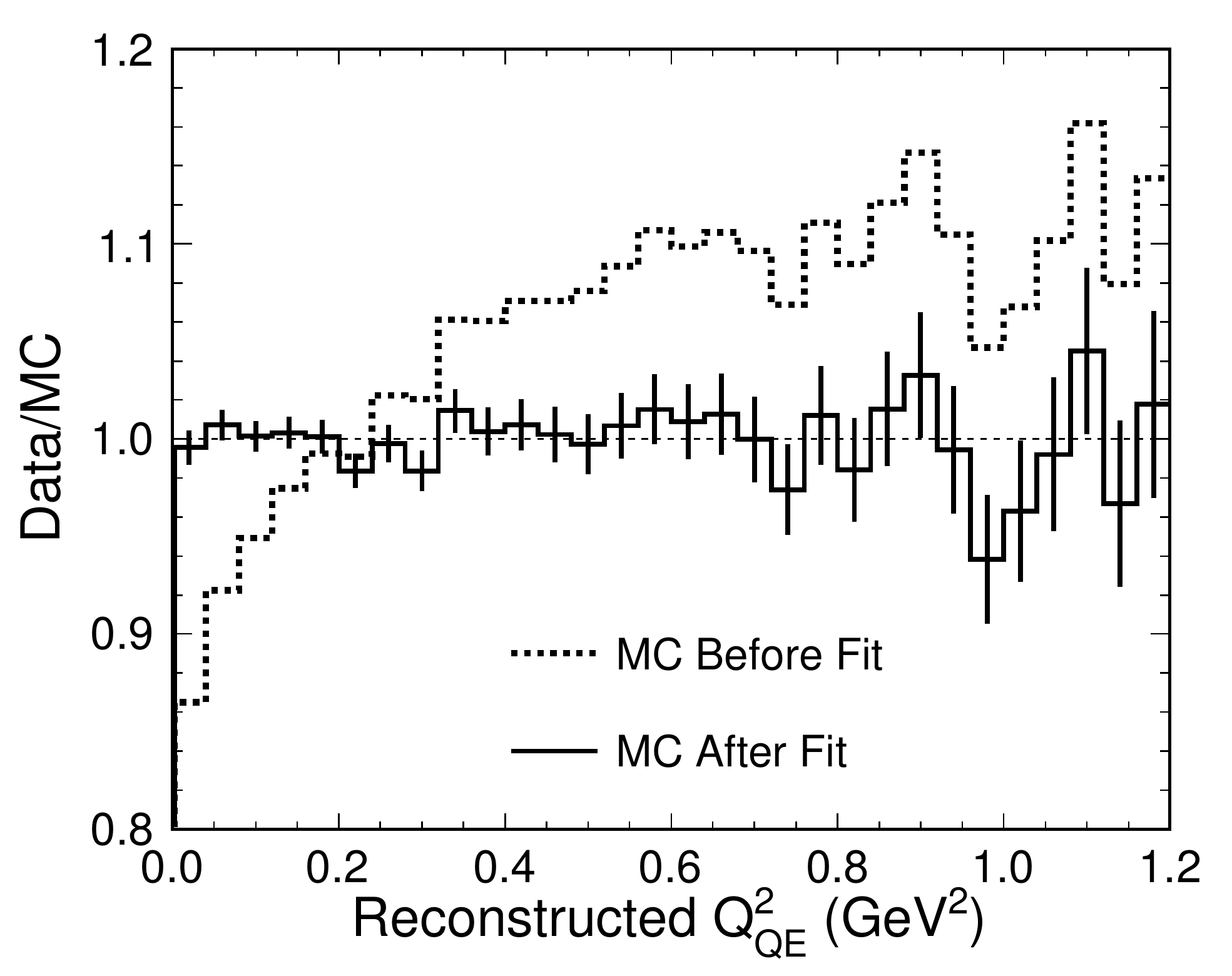}
\caption{MC predictions (dotted, solid-line histograms) compared to the distribution 
of reconstructed $Q^{2}_{QE}$ for events of the CCQE enhanced sample (solid circles).
The plots show that the best fit MC, which is fitted to the shape of the data distribution 
over the range of allowed $Q^{2}_{QE}$, agrees very well
with the data (top, solid-line vs data points).   For the best-fit MC, the 
data-over-MC ratio (bottom, solid-line histogram) equals
1.0 to within $\pm 4\%$ in all bins
over the full Q$^{2}$ range.
\label{fig:fitResultsFig2}}
\end{center}
\end{figure}

Comparisons of the default (dotted-line histogram) 
and best-fit (solid-line histogram) MC $Q^{2}_{QE}$ spectra to the data distribution
are presented in Fig.~\ref{fig:fitResultsFig2}.   
The matchup of distributions is shown in the upper plot,  
and the ratio of the data to the predicted MC distribution is displayed in the
lower plot.  
In the upper plot,  the principal fit 
(upper row of Table~\ref{tab:fitResultsTab})  
is seen to provide an excellent description of the data distribution 
over the full range of $Q^{2}_{QE}$ considered by this analysis.    This is not
the case for the original reference MC.   
As is apparent in Fig.~\ref{fig:fitResultsFig2}\,(top) and is made explicit by the ratio displayed in 
Fig.~\ref{fig:fitResultsFig2}\,(bottom),  the shape predicted 
by the reference MC describes a spectrum that lies above
the data for $Q^{2} < 0.15$\,GeV$^{2}$ 
and falls below the data for $Q^{2} > 0.25$\,GeV$^{2}$.

\subsection{Systematic errors for determination of $M_{A}$}
\label{subsec:Systematic-errors}

The principal fit treats three sources of uncertainty using nuisance parameters whose values {\it ab initio}
are known to within certain ranges.   There are, however, systematic uncertainties
whose contribution cannot be captured in that way.   These include errors inherent 
to the event reconstruction, to the analysis procedure, and to model uncertainties.
The approach taken here is to set the relevant event selection or model parameter 
to its $\pm 1\sigma$ values and then to refit the MC to the CCQE enhanced data sample.
The deviations ($\pm$) in best-fit $M_{A}$ which result from the variations comprise the error
estimate to the $M_{A}$ determination arising from a particular systematic.
There are eight sources of systematic error whose individual contribution to the error budget is 
comparable to statistical fluctuations.  Their identity and evaluation are
described below, in descending order of estimated error contribution:

\begin{enumerate}
  \item { \it Intranuclear scattering of produced hadrons}:   Pions and nucleons produced 
  in the initial $\nu$N interactions in iron can re-interact within the nucleus before emerging
  to produce the observed final state.   Their alteration of final states is accounted for by {\small NEUGEN3} using 
  an intranuclear cascade model.   The model contains parameters which govern the
  effective cross sections for pion and nucleon rescattering.    The parameters are set
  according to published data on $\nu$N, $\gamma$N, and $\pi$N scattering, however
 each parameter has an error range.     Changes in these parameters cause event migrations in the MC
 across the $E_{had}$ selection that defines the CCQE sample~\cite{minos-nucl-effects}.   
 In a detector where a large fraction of the target
 material is passive, events are moved into and 
 out of the $E_{had} \simeq 0$ region (see Fig.~\ref{fig:ccSample_eshw}).
 
 \smallskip
 Trial fits were carried out in which the MC was reweighted to simulate a $\pm$1$\sigma$
 change to an intranuclear scattering parameter~\cite{gallag-gen-val};  separate trials were carried out for each of the
ten parameters.   For all trials the best-fit values for $M_{A}$ and for the nuisance parameters were observed to remain within 
the 1$\sigma$ error range of the nominal value.   The main uncertainties are associated with 
nucleon absorption, pion absorption, and with the hadron formation time. 
(The formation time determines the point at which a produced hadron acquires 
its full scattering cross section.)   Uncertainties with other parameters are either negligible or 
strongly correlated with the errors of these three processes.   The quadrature sum
from the individual (maximum magnitude) variations to $M_{A}$ induced by parameter
variations for these three intranuclear rescattering processes is taken as the overall 
error estimate listed in the first row of Table \ref{tab:systematicsTab}.
\item {\it CCQE selection using visible hadronic energy}:  The primary CCQE selection
requires the reconstructed energy of the hadronic system recoiling from the muon to
be less than 225 MeV.   This cut removes most
CC DIS events since these processes tend to have relatively large $E_{had}$ values. 
It also reduces the amount of CC baryon resonance production which remains as a background.
The cut value chosen lies at the midpoint of an $E_{had}$ range 
characterized by small and regular changes in sample purity 
and in MC-vs-data discrepancy with incremental variation in $E_{had}$.  
This region of relative stability extends for $\pm$75 MeV on either side 
of the designated cut value at 225 MeV.    The uncertainty inherent to placement 
of the cut is evaluated using a set of trial fits of the MC to data in which the assigned $E_{had}$ cut  value
 is stepped from 150 to 300 MeV.   The maximum variation in the fit outcomes provides the error estimate
 (2nd row of Table \ref{tab:systematicsTab}).
\item{ \it Uncertainties in detector modeling}:
The detector is divided longitudinally ($z$-coordinate) into calorimeter 
and spectrometer sections.   The analysis fiducial volume
is located asymmetrically with respect to the detector's transverse, 
horizontal dimension ($x$-coordinate) and with respect to the
toroidal magnetic field.    In trial fits using subsamples selected 
from different regions of the fiducial volume, small 
shifts of fit parameter values are observed which correlate 
with event vertex location.  These shifts have a non-statistical
component and appear to be associated with uncertainties in detector modeling.  
Their presence implies a systematic uncertainty
for the $M_A$ measurement.    An error estimate is obtained 
by relating excursions observed in $E_{\mu}$ and $\theta_{\mu}$
to corresponding trends in $Q^2$, and then evaluating the variation 
that propagates to the $M_A$ determination.   
Excursions of potential significance are observed with sample splitting 
based upon vertex $z$ or upon vertex $x$.  On the other hand
negligible variations are found when splitting the sample according to 
vertex $y$;  the distribution of event vertices exhibits vertical symmetry in the fiducial region, as expected. 
The excursions associated with sample subdivisions
using vertex $z$ or vertex $x$  imply shifts propagated to $M_A$ of 4.0\% and 3.3\% respectively.    
The uncertainty assigned to the best-fit value of the $M_A$ measurement is taken as the quadrature sum.  
  \item {\it Low-Q$^{2}$ suppression of baryon resonance production}:  A suppression weight
  has been added to the MC modeling of CC baryon resonance production at low $Q^{2}$,
  as described in Sec.\,\ref{subsec:Suppression}.   
   A systematic error is assigned to the utilization of this weight.   It represents uncertainties associated with the
  shape of the weight function, in particular with its representation of the approach to null suppression 
  as a function of increasing $Q^2$.   The error
  is estimated by shifting the suppression function in accordance with its $\pm$1$\sigma$ error band
  and then re-fitting to find the resulting variation in $M_{A}$ (4th row, Table \ref{tab:systematicsTab}).
\item {\it Hadronic energy MC-vs-data offset}:  Discrepancies may exist 
between energies assigned to visible hadronic activity in data versus 
the MC at the level of tens of MeV.    Sources include offsets in the 
calorimeter response to stopping pions and/or protons and data-vs-MC difference 
in the effect of nucleon binding energy on reconstructed $E_{had}$.   
Such offsets cause a small migration of MC events across the 
$E_{had}$ selection boundary.   For the above-mentioned sources, 
an upper bound of 20 MeV is estimated
for the magnitude of the net offset.    On the basis of trial fits in which the 
$E_{had}$ cut for the MC was varied by $\pm 20$ MeV, the uncertainty propagated to $M_A$ was ascertained. 
\item {\it CC DIS cross section}:  Approximately 11\% of the CCQE enhanced sample consists of CC DIS events.
In the MC, the DIS cross section for scattering into low-multiplicity pion production
channels is implemented by a combination of KNO and Bodek-Yang~\cite{BodekYang:2004} models.   The relative 
cross section rates, among CC channel combinations of target nucleon with multiple charged and neutral pions,
are governed by a parameter set which, upon introduction of isospin constraints, reduces to four parameters.
Uncertainty ranges are assigned to these parameters by {\small NEUGEN3} on the basis of limited knowledge of the cross sections.
The sensitivity of DIS contributions to the sideband samples is not sufficient to further constrain these errors.
Fit trials were conducted in which the parameters were varied individually over their $\pm 1\sigma$ ranges
and the fit to $M_{A}$ was redone.   The maximum $M_{A}$ displacement
for each parameter was added in quadrature to obtain the estimated systematic error. 
The dominant contribution to this error arises from cross section uncertainties with CC two-pion channels.
\item{ \it Correction to muon angular resolution in the MC}:
As described in Sec.~\ref{subsec:Muon-Ang-Res}, a $Q^2_{QE}$-dependent weight
 is applied to MC events of the CCQE enhanced sample to ensure 
 that muon angular resolution of the MC represents the resolution observed in the data.   
 The required correction is found to be nearly identical across all subsamples of the analysis; 
 the method of correction is insensitive to the underlying $M_{A}$ in a sample. 
The determination of MC-vs-data resolution difference per bin of muon track length has uncertainties, and these define the
error envelope associated with the correction weight applied to MC events.   The one-sigma variations allowed by the envelope impart
an uncertainty to the $M_A$ determination of  the amount shown in row 7 of Table~\ref{tab:systematicsTab}. 
\item {\it NuMI flux-tuning parameters}:  The NuMI flux calculation used by the MC includes tuned parameter
settings that characterize the beam optics and the production of hadrons from the primary target~\cite{prd_1e20}.
Changes of $\pm$1$\sigma$ are considered for each beam-optics parameter; also considered are the 
differences between calculated versus data-tuned settings for the hadro-production parameters.   By design, the analysis
is insensitive to the absolute scale of the neutrino flux, and the distribution shape for $Q^{2}_{QE}$ is also
fairly insensitive to uncertainties in the spectral shape of the neutrino flux.    Consequently these flux uncertainties
give a sub-percent contribution to the systematic error budget. 
\end{enumerate}

\begin{table}
\begin{tabular}{c|cc}
\hline
\hline
 & \multicolumn{2}{c}{ \rule{0pt}{2.5ex} Fit $Q^{2}$ Range} \\ 
Systematic Error & \multicolumn{2}{c}{ \rule{0pt}{2.1ex} $0.0 < Q^{2}_{QE} < 1.2$ GeV$^{2}$} \\
\cline{2-3}
Source & (+) Shift  & (-) Shift  \\
& (GeV) & (GeV) \\
\cline{1-3}
Intranuclear scattering  & ~0.066 & ~0.066\\
\cline{1-3}
CCQE $E_{had}$ selection  & ~0.062 & ~0.062\\
\cline{1-3}
Detector model in $x$, $z$ & ~0.059 & ~0.059\\
\cline{1-3}
$\Delta$/N$^{*}$ low-$Q^2$ suppression & ~0.005 & ~0.088\\
\cline{1-3}
Hadronic energy offset & ~0.047 & ~0.045\\
\cline{1-3}
DIS cross section & ~0.024 & ~0.022\\ 
\cline{1-3}
$\mu^{-}$ angular resolution & ~0.016 & ~0.015\\
\cline{1-3}
Flux tuning parameters& ~0.008 & ~0.008\\
\cline{1-3}
 & & \\
Total Syst. Error (GeV) & + 0.122 & - 0.149\\
\hline
\hline
\end{tabular}
\caption{
Shift from nominal in the value of $M_{A}$ resulting 
from variation of each systematic error source.   The systematics listed here are evaluated
separately from the nuisance parameters of the principal fit.
\label{tab:systematicsTab}}
\end{table}

Referring to Table \ref{tab:systematicsTab}:   For each error source (left-hand column), the
shifts in the axial-vector mass from the best-fit nominal value 
(lower row of Table~\ref{tab:fitResultsTab}) are presented in the second and third
columns of Table \ref{tab:systematicsTab} for systematic parameter variations of $+ 1\sigma$ and $- 1\sigma$ respectively.  
The bottom row displays the quadrature sums of the systematic errors.
The sums represent the systematic error contribution arising from all sources other than those treated by the nuisance parameters of the principal fit.

 The QE-enhanced data sample is essentially 
 a CC single-track sample, and the requirement that the
 detected hadronic system be of zero or low energy is central to the event selection.    
 Table~\ref{tab:systematicsTab} shows that uncertainties associated with the selection of
 $E_{had}$ (rows 2 and 5) contribute significantly to the systematic error.   These errors, together with 
 the contribution from intranuclear rescattering (row 1), are intrinsic to the use of thick iron plates
 in the detector.   Similarly, the uncertainties associated with detector modeling arise from 
 asymmetries in the detector configuration and are not amenable to significant further reduction.

  
\section{Results and Discussion}
\label{sec:Results-Discuss}

Charged current $\numu$Fe interactions initiated by a broad-band neutrino flux 
peaked at $\sim$ 3.0 GeV are examined with high statistics using a three-stage analysis.  
In the first stage, final states are examined inclusively using distributions 
in visible hadronic energy, neutrino energy, $Q^2$, and in hadronic mass $W$.  A conventional MC model 
using an RFG nucleus and with CCQE scattering, baryon-resonance production, and 
inelastic scattering/DIS as the predominant interaction categories, 
is found to give rough but respectable characterizations of the data 
(Figs.~\ref{fig:ccSample_eshw}, \ref{fig.ccSampleEnuAndQ2}, and~\ref{fig.CC-Inc-W-distribution}).   

\smallskip

This characterization guides the second stage in which the CC inclusive sample is
broken out into independent subsamples, each containing a distinctive
mixture of the three main reaction categories.   One subsample, selected to be
enriched in CCQE events, is put aside for the third stage.   The remaining
four subsamples are dominated by baryon-resonance and inelastic/DIS events.
The shapes of data $Q^2$ distributions for the latter samples are then compared to
the MC model (Figs.~\ref{fig.dissamples}, \ref{fig.resdistransample}, and~\ref{fig.qeressamples}).   

Here also the MC manages respectable description
by and large, however at low-$Q^2$ its predictions exceed the data in subsamples containing
sizable amounts of baryon-resonance production (of mostly $\Delta(1232)$ states).
 Motivated by this correlation, 
and with knowledge of the evidence given by MiniBooNE for baryon-resonance suppression 
in neutrino-carbon interactions at low-$Q^2$~\cite{MiniBooNEMA:2010, MiniBooNEPiplus:2011, MiniBooNEPizero:2011}, 
a suppression function is developed whose $Q^2$ dependence 
is displayed in Fig. 9.    The analysis incorporates 
this low-$Q^2$ suppression of baryon resonances into its otherwise conventional 
MC treatment of neutrino-nucleus scattering 
for the purpose of fitting the CCQE-enriched subsample.

\smallskip

The CCQE enhanced subsample is the focus of the analysis third stage.  
Its distribution in $Q^{2}_{QE}$, for neutrinos 
of $1.0 < E_{\nu} < 8.0$\,GeV, is presented in Fig.\,\ref{fig:fitResultsFig2}.
This sample contains 221,297 events of which 66\% are estimated 
to be quasielastic interactions (see Table~\ref{tab.SelectedEvents}).
The shape of the $Q^{2}_{QE}$ distribution of the CCQE-enriched data sample
is fitted using a $\chi^2$ in which the axial-vector mass $M_{A}$ is a free parameter, and 
the muon energy scale, the axial-vector mass for baryon resonance production, and an 
effective low-$Q^{2}_{QE}$ suppression are treated using nuisance parameters.
For the effective $M_{A}$ value which sets the $Q^2$ scale in the empirical dipole 
axial-vector form factor of neutrons bound within iron nuclei,  the best-fit value is:
\begin{equation}
\label{eq:M_A-final-result}
M_{A} = 1.23 ^{+0.13} _{-0.09} \mbox{(fit)} ^{+0.12} _{-0.15} \mbox{(syst.)}~\mbox{GeV}.
\end{equation}
The mean neutrino energy for the fitted signal sample is $\langle E_{\nu} \rangle = 2.79$\,GeV.
The error range obtained by the fit includes
the effects of finite sample statistics plus the variations 
and correlations allowed by the nuisance parameters.   
The uncertainty introduced by the systematic error sources 
is additional to that which is estimated by the fit.   It
is listed separately in Eq.~\eqref{eq:M_A-final-result}.    

\smallskip
 
The best-fit MC result, as shown in Fig.~\ref{fig:fitResultsFig2}, gives 
an excellent description of the shape of the data $Q^{2}_{QE}$ distribution 
over the range~$0.0 < \,$$Q^{2}_{QE} < 1.2$\,GeV$^{2}$.
Compared to the original MC reference model,
 the data prefers a $Q^2$ spectrum which is harder (flatter) through
this range.   As shown in Fig.~\ref{fig:lowQ2Suppression}, the data also prefers   
that a small amount of additional rate suppression be added at low $Q^{2}$.

\smallskip

As related in Sec.~\ref{sec:CCQE}, the axial-vector mass of CCQE scattering on free nucleons is
generally regarded to be $\sim$ 30-40 MeV lower than 
the 2002 compilation value (1.026 $\pm~0.021$)\,GeV of Ref.~\cite{Bernard:2002}.
The effective $M_{A}$ value for CCQE interactions in iron nuclei determined by this analysis 
lies above the free-nucleon value, although with allowance for systematic uncertainty the
disagreement is only at the level of 1.4 $\sigma$.   Table~\ref{tab:systematicsTab} shows that 
no single source dominates the systematic error assigned to $M_{A}$, hence further reduction of the total error
would be difficult to accomplish with MINOS data.    
Among the five leading systematics there are four
arising from the Near Detector which was originally designed for measurement of 
$\numu$ disappearance due to oscillations.

The MINOS effective $M_{A}$ value is in agreement
with the K2K result for interactions in oxygen:  $M_{A}$ = (1.20 $\pm~0.12$)\,GeV~\cite{K2KMA:2006}.   
It is also compatible with the relatively high nominal value obtained by MiniBooNE for CCQE interactions
on carbon: $M_{A}$ = (1.35 $\pm~0.17$)\,GeV~\cite{MiniBooNEMA:2010}.  
Notable perhaps with the MINOS result is the absence of an upward trend in effective $M_{A}$ when a distinctly
larger target nucleus is used.   The MINOS value, together with the K2K and MiniBooNE results,
are consistent with interpretations~\cite{Martini:2009uj, Gallagher-ARNS, Morfin:2012kn, Gran:2013kda}
that large values for the effective $M_{A}$ reflect nuclear medium effects not accounted for using the 
Fermi gas treatment of the nucleus.   On the other hand, 
the value presented in Eq.~\eqref{eq:M_A-final-result} lies above the NOMAD measurement
for high energy $\numu$--carbon scattering:  $M_{A}$ = (1.05 $\pm~0.06$)\,GeV~\cite{NOMADMA:2009}.
Their result is based upon a combined sample of 1-track and 2-track events with 
$3 < E_{\nu} < 100$ GeV and with $Q^2$ extending to 2.0 GeV$^2$.  

\smallskip

In summary, an investigation of CC $\numu$ interactions on iron is reported which bridges the
neutrino energy ranges previously examined by experiments using light-nucleus targets.
Event distributions in kinematic variables are presented for CC inclusive scattering, and
for subsamples selected to have distinctly different populations of CCQE, baryon resonance
production, and inelastic/DIS events.   For all distributions, comparisons are given to 
predictions of an MC simulation based upon conventional phenomenology with
neutrinos interacting with quasi-free nucleons in a nucleus modeled as a relativistic Fermi gas.
From these comparisons it is inferred that
CC baryon resonance production, the principal background to CCQE, is subject to
a $Q^2$-dependent suppression of rate 
with the functional form shown in Fig.~\ref{fig.errorband}.
With inclusion of this suppression effect into the MC simulation, 
the $Q^2$ distribution for CCQE scattering in iron is found to be well-described using an effective axial-vector mass
with the value given in Eq.~\eqref{eq:M_A-final-result}.  

 These results provide new information 
for development of more realistic models of charged-current neutrino-nucleus scattering and of
nuclear medium effects at work in CCQE and CC baryon resonance production.   Improved models are needed as 
benchmarks for interpreting neutrino scattering data and as guides to precise determinations of the atmospheric
mixing angle and of other neutrino oscillation parameters~\cite{Huber-2013}.


\begin{acknowledgments}
This work was supported by the U.S. DOE; the U.K. STFC; the U.S. NSF; the
State and University of Minnesota; the University of Athens, Greece;
and Brazil's FAPESP, CAPES, and CNPq.  We thank the staff of Fermilab 
for their invaluable contributions to the research of this work.
\end{acknowledgments}


\end{document}